\documentclass[useAMS,usenatbib]{mn2e}

\usepackage{graphicx}
\usepackage[draft]{hyperref}
\usepackage{ifthen}
\usepackage{amsmath}
\usepackage{amssymb}

\def\d{{\rm d}}

\newcounter{CDMDone}
\def\CDM{\ifthenelse{\equal{\arabic{CDMDone}}{0}}{cold dark matter (CDM)\setcounter{CDMDone}{1}}{CDM}}

\newcounter{WDMDone}
\def\WDM{\ifthenelse{\equal{\arabic{WDMDone}}{0}}{warm dark matter (WDM)\setcounter{WDMDone}{1}}{WDM}}

\title[Dark Matter Halo Merger Histories Beyond Cold Dark Matter]{Dark Matter Halo Merger Histories Beyond Cold Dark Matter: I - Methods and Application to Warm Dark Matter}
\author[Benson, Farahi, Cole, Moustakas, Jenkins, Lovell, Kennedy, Helly \& Frenk]{Andrew J. Benson$^1$\thanks{E-mail: {\tt abenson@obs.carnegiescience.edu}}, Arya Farahi$^2$, Shaun Cole$^3$, Leonidas A. Moustakas$^4$, \newauthor
Adrian Jenkins$^3$, Mark Lovell$^3$, Rachel Kennedy$^3$, John Helly$^3$ \& Carlos Frenk$^3$\\
$^1$ Carnegie Observatories, 813 Santa Barbara Street, Pasadena, CA 91101, U.S.A. \\
$^2$ Michigan Center for Theoretical Physics, Randall Laboratory of Physics, The University of Michigan, Ann Arbor, MI 48109-1120, USA \\
$^3$ Institute for Computational Cosmology, Department of Physics, University of Durham, South Road, Durham DH1 3LE \\
$^4$ Jet Propulsion Laboratory, California Institute of Technology, 4800 Oak Grove Dr., Pasadena, CA 91109, USA \\
}

\begin{document}

\maketitle

\begin{abstract}
We describe a methodology to accurately compute halo mass functions, progenitor mass functions, merger rates and merger trees in non-cold dark matter universes using a self-consistent treatment of the generalized extended Press-Schechter formalism. Our approach permits rapid exploration of the subhalo population of galactic halos in dark matter models with a variety of different particle properties or universes with rolling, truncated, or more complicated power spectra. We make detailed comparisons of analytically derived mass functions and merger histories with recent warm dark matter cosmological N-body simulations, and find excellent agreement. We show that, once the accretion of smoothly distributed matter is accounted for, coarse-grained statistics such as the mass accretion history of halos can be almost indistinguishable between cold and warm dark matter cases. However, the halo mass function and progenitor mass functions differ significantly, with the warm dark matter cases being strongly suppressed below the free-streaming scale of the dark matter. We demonstrate the importance of using the correct solution for the excursion set barrier first-crossing distribution in warm dark matter--if the solution for a flat barrier is used instead the truncation of the halo mass function is much slower, leading to an overestimate of the number of low mass halos.
\end{abstract}


\section{Introduction}

The \CDM\ paradigm \citep{blumenthal_formation_1984} works extremely well on large scales \citep{seljak_cosmological_2005,percival_shape_2007,ferramacho_constraints_2009,sanchez_cosmological_2009,komatsu_seven-year_2011}, but there remains the possibility of deviations from the \CDM\ expectations on small scales--arising notably from the issue of cores vs. cusps and the inner slope of dark matter density profiles \citep{salucci_constant-density_2001,donato_cores_2004,newman_distribution_2009,donato_constant_2009,de_blok_core-cusp_2010,kuzio_de_naray_recovering_2011,kuzio_de_naray_baryons_2011,newman_dark_2011,salucci_dwarf_2012,wolf_dark_2012} and the apparent paucity of bright satellites around the Milky Way \citep{boylan-kolchin_too_2011,boylan-kolchin_milky_2012}. Several extensions to dark matter phenomenology  \citep{markevitch_constraints_2004,boehm_constraints_2005,ahn_formation_2005,miranda_constraining_2007,randall_constraints_2008,boyarsky_lyman-_2009,lovell_haloes_2012} and several different particle physics candidates for dark matter \citep{raffelt_astrophysical_1990,turner_windows_1990,jungman_supersymmetric_1996,hogan_new_2000,spergel_observational_2000,abazajian_sterile_2001,cheng_kaluza-klein_2002,feng_superweakly_2003,sigurdson_charged-particle_2004,hubisz_phenomenology_2005,feng_hidden_2009} have been put forward to explain these deviations, although it remains unclear if any of these proposals is able to fully explain observed phenomena \citep{kuzio_de_naray_case_2010} or if they are even necessary, with the observed phenomena simply being a consequence of galaxy formation physics in a \CDM\ universe \citep{benson_effects_2002,libeskind_satellite_2007,li_common_2009,stringer_physical_2010,font_population_2011,oh_central_2011,governato_cuspy_2012,kuhlen_dwarf_2012,pfrommer_cosmological_2012,pontzen_how_2012,starkenburg_satellites_2012,vera-ciro_not_2012}. A variety of experimental measurements may be sensitive to the small-scale structure of dark matter halos \citep{simon_high-resolution_2005,viel_how_2008}. The most promising are future lensing experiments, which have the potential to strongly constrain dark matter particle phenomenology \citep{keeton_new_2009,vegetti_bayesian_2009,vegetti_statistics_2009,vegetti_quantifying_2010,vegetti_detection_2010,vegetti_gravitational_2012}. A variety of previous work has shown that the subhalo mass function should depend sensitively on the dark matter physics and the shape of the power spectrum. To maximize the scientific return of future experiments therefore requires the ability to accurately and rapidly predict the distribution of dark matter substructure as a function of dark matter particle thermal or interaction properties, for arbitrary power spectra.

In this work, we develop techniques to follow the growth of nonlinear structure in non-\CDM\ universes using a fully consistent treatment of the extended Press-Schechter formalism. We demonstrate the performance of these techniques by applying them to a representative case of \WDM, which has the advantage of several pre-existing N-body simulations which we utilize to test the accuracy of our methods. \WDM\ particles\footnote{The two usual candidates--both lying beyond the standard model of particle physics--are sterile neutrinos \protect\citep{dodelson_sterile_1994,shaposhnikov_msm_2006} and gravitinos \protect\citep{ellis_cosmological_1984,moroi_cosmological_1993,kawasaki_gravitino_1997,gorbunov_is_2008}.} are lighter than their \CDM\ counterparts, allowing them to remain relativistic for longer in the early universe and to retain a non-negligible thermal velocity dispersion. This velocity dispersion allows them to free-stream out of density perturbations and so suppresses the growth of structure on small scales \citep{bond_collisionless_1983,bardeen_statistics_1986}. While mass functions have been previously considered in this case \citep{barkana_constraints_2001}, we go one step further and develop methods to compute conditional mass 
functions, and halo merger rates and use these to construct merger trees in \WDM\ universes. These merger trees are a key ingredient required to predict the distribution of substructure masses, positions and internal structure as is necessary to make detailed predictions for future lensing experiments.

The remainder of this paper is organized as follows. In \S\ref{sec:methods} we describe the changes that we introduce to the extended Press-Schechter formalism to make it applicable to the case of non-\CDM\ scenarios (including some specifics for the \WDM\ case). In \S\ref{sec:wdmResults} we apply these methods to the case of \WDM, first comparing their predictions to the available data from N-body simulations, then exploring the limitations of ignoring the effects of \WDM\ velocity dispersion, and presenting a comparison of key results between \WDM\ and \CDM. Finally, in \S\ref{sec:discuss} we discuss the consequences of this work and present our conclusions.

We also include two appendices. Appendix~\ref{app:numericalMethod} gives a detailed derivation of our numerical procedure for solving the excursion set first crossing problem for arbitrary barriers. Appendix~\ref{app:numerics} explores the numerical accuracy and robustness of the methods developed in this work.

When comparing our analytic theory with results from N-body simulations we will adopt the same cosmological parameters and dark matter particle properties as were used for the simulation. These values will be listed where relevant. For the rest of this work, specifically in \S\ref{sec:methods}, \S\ref{sec:limits}, \S\ref{sec:wdmVsCdm} and Appendix~\ref{app:numerics} we adopt a canonical a cosmological model with $\Omega_{\rm M}=0.2725$, $\Omega_\Lambda=0.7275$, $\Omega_{\rm b}=0.0455$ and $H_0=70.2$~km s$^{-1}$ Mpc$^{-1}$
\citep{komatsu_seven-year_2011} and a canonical \WDM\ particle of mass, $m_{\rm X}=1.5$~keV and effective number of degrees of freedom $g_{\rm X}=1.5$ (the expected value for a fermionic spin-$\frac{1}{2}$ particle).

\section{Methods}\label{sec:methods}

Our approach makes use of the Press-Schechter formalism \citep{press_formation_1974,bond_excursion_1991,bower_evolution_1991,lacey_merger_1993} which, after substantial development and tuning against N-body simulations, has proven to be extremely valuable in understanding the statistical properties of dark matter halo growth in \CDM\ universes. The Press-Schechter formalism in its modern form is expressed in terms of excursion sets--the set of all possible random walks in density at a point as the density field is smoothed on ever smaller scales. Halo formation corresponds to a random walk making its first crossing of a barrier. The height of that barrier is determined from models of the non-linear collapse of simple overdensities.

The Press-Schechter algorithm requires three ingredients: 1) the power spectrum of fluctuations in the density field (characterized by $\sigma(M)$, the fractional root-variance in the linear-theory density field at $z=0$); 2) the critical threshold in linear-theory corresponding to the gravitational collapse of a density perturbation, $\delta_{\rm c}$; and 3) a solution for the statistics of excursion sets to cross this threshold. We will address each of these three ingredients below.

\subsection{Power Spectrum}

We assume a power-law primordial power spectrum with $n_{\rm s}=0.961$ \citep{komatsu_seven-year_2011}, and adopt the transfer function of \cite{eisenstein_power_1999}. We include a modification for warm dark matter using the fitting function of \citeauthor{bode_halo_2001}~(\citeyear{bode_halo_2001}; as re-expressed by \citealt{barkana_constraints_2001}) to impose a cut-off below a specified length scale, $\lambda_{\rm s}$:
\begin{equation}
T(k) \rightarrow T(k) \left[1 + (\epsilon k \lambda_{\rm s})^{2 \nu}\right]^{-\eta/\nu},
\label{eq:transferFunction}
\end{equation}
where $\epsilon=0.361$, $\eta=5$ and $\nu=1.2$ are parameters of the fitting function. For our canonical \WDM\ particle, the smoothing scale\footnote{This scale is usually approximated as being equal to the speed of the particles at the epoch of matter-radiation equality multiplied by the comoving horizon scale at that time; see \protect\cite{bode_halo_2001} for further discussion.} is $\lambda_{\rm s}=0.124$~Mpc (\citealt{barkana_constraints_2001};~eqn.~4) corresponding to a mass of $M_{\rm s}=4 \pi \bar{\rho}\lambda_{\rm s}^3/3=2.97\times 10^8$~M$_\odot$. The power spectrum is normalized to give the required $\sigma_8=0.807$ \citep{komatsu_seven-year_2011} when integrated under a real-space top-hat filter of radius $8h^{-1}$Mpc (where $h=H_0/100$~km s$^{-1}$ Mpc$^{-1}$).

\subsubsection{Window Function}\label{sec:windowFunction}

To derive the variance, $S(M)$, or, equivalently, the root-variance, $\sigma(M)\equiv\sqrt{S(M)}$, from the power spectrum a window function, $W(k|M)$, must be adopted. Specifically,
\begin{equation}
 S(M)\equiv \sigma^2(M) = {1 \over 2\pi^2} \int_0^\infty 4 \pi k^2 P(k) W^2(k|M) {\rm d} k.
\end{equation}
In the \cite{bond_excursion_1991} re-derivation of the \cite{press_formation_1974} mass function and 
extended Press-Schechter conditional mass functions one assumes a sharp 
$k$-space filtering,
\begin{equation}
 W(k|M) = \left\{ \begin{array}{ll} 1 & \hbox{if } k \le k_{\rm s}(M) \\ 0 & \hbox{if } k > k_{\rm s}(M), \end{array}\right.
 \label{eq:ksharp}
\end{equation}
of the linear density field so that the trajectories 
of density fluctuation, $\delta$, versus mass scale are true Brownian 
random walks. Despite the use of this sharp 
$k$-space filtering in deriving the extended Press-Schechter solutions, it has been conventional to adopt the $\sigma(M)$ that is 
given by a real-space top-hat window function,
\begin{equation}
 W(k|M) = {3 [\sin(kR)-kR \cos(kR)] \over (kR)^3},
 \label{eq:tophat}
\end{equation}
when utilizing those solutions. The reason for this is 
that only in
this case is there no ambiguity between the filtering scale, $R$, and 
the corresponding mass,
\begin{equation}
 R=\left({3 M \over 4 \pi \bar{\rho}}\right)^{1/3}
 \label{eq:rtophat}
\end{equation}
(see \citealt{lacey_merger_1993}).  
For \CDM\ power spectra, which are essentially pseudo power-laws with a 
slowly varying slope, the shape of the $\sigma(M)$ function varies little 
with the choice of filter function.

However, in the case of power 
spectra with a small scale cut-off, such as those of \WDM\ models and many other non-\CDM\ candidates, the 
situation is very different. With a top-hat window function, $\sigma(M)$ 
continues to increase with decreasing filter mass even for masses well 
below the cutoff scale.  This happens because although there are no new 
intrinsic small scale modes
entering the filter, the longer wavelength modes are getting re-weighted 
as the mass scale of the filter increases.
In contrast, with a sharp-$k$ filter, $\sigma(M)$ increases monotonically 
up to the cutoff scale and then becomes
constant with the transition being determined by the abruptness of the 
cutoff in the input \WDM\ power spectrum.
Since the Press-Schechter mass functions depend directly on $\sigma(M)$, 
including a direct dependence on the
logarithmic slope of $\sigma(M)$, its predictions for the case of \WDM\ 
are sensitive to this choice. Hence for non-\CDM\ models
it is important to revisit the issue of filter choice and one would 
expect that the conventional top-hat choice will
lead to an overestimate of the number of low mass haloes (an effect apparent in the works of \cite{barkana_constraints_2001} and \cite{menci_galaxy_2012}).

We begin by adopting the sharp $k$-space window function of  eqn.~(\ref{eq:ksharp}) when working with truncated power spectra, such that the contribution from long-wavelength modes remains constant. In the case of a top-hat real-space window function there is a natural relation between the top-hat radius and the smoothing mass scale, as given by eqn.~(\ref{eq:rtophat}). In the case of a sharp $k$-space window function, the choice of $k_{\rm s}(M)$ is less clear. \cite{lacey_merger_1993} suggest choosing $\stackrel{\sim}{W}\!\!(r=0)=1$ (where $\stackrel{\sim}{W}\!\!(r)$ is the Fourier transform of the window function) and then choosing $k_{\rm s}(M)$ such that the integral of the mean density, $\bar{\rho}$, under $\stackrel{\sim}{W}\!\!(r)$ equals the required smoothing mass, $M$. However, as noted by \cite{lacey_merger_1993}, this choice for $k_{\rm s}(M)$ lacks any strong physical motivation.

Therefore, we advocate a different approach, namely we choose
\begin{equation}
k_{\rm s} = a / R,
\label{eq:kToR}
\end{equation}
where $R$ is computed from eqn.~(\ref{eq:rtophat}) and the parameter $a$ is chosen such that the turnover in the halo mass function occurs in the same location as that seen in N-body simulations of \WDM, as will be shown in \S\ref{sec:Nbody}. We find that $a=2.5$ is required to meet this condition\footnote{The normalization advocated by \protect\cite{lacey_merger_1993} is equivalent to $a=(9\pi/2)^{1/3}\approx2.42$ which is not too different from our best-fit value of $2.5$}.. This will mean that $\sigma(M)$ differs from the usual result on large scales, resulting in a difference in the mass function for high mass halos that would not be expected. To remedy this, we note that the critical overdensity for collapse is usually motivated on the basis of spherical top-hat collapse models. Since we are no longer using a top-hat filter we choose to allow freedom in the choice of the critical overdensity such that the halo mass function is unchanged for high masses. This point will be discussed in greater detail in \S\ref{sec:barrierScaling}.

We note that fixing the $k_{\rm s}(M)$ relation in this way introduces a free parameter, $a$, into our method and introduces a dependence on the N-body simulation to which we will compare our results in \S\ref{sec:NbodyMassFunction}. However, given that the \CDM\ linear power spectrum is close to a power-law on the scales that will be of interest for \WDM\ models, and that the modification to the transfer function due to \WDM\ scales simply with the mass of the dark matter particle we expect that the same choice of $a$ should be valid for all \WDM\ particle masses of interest.

\begin{figure}
\begin{center}
\includegraphics[width=8.5cm]{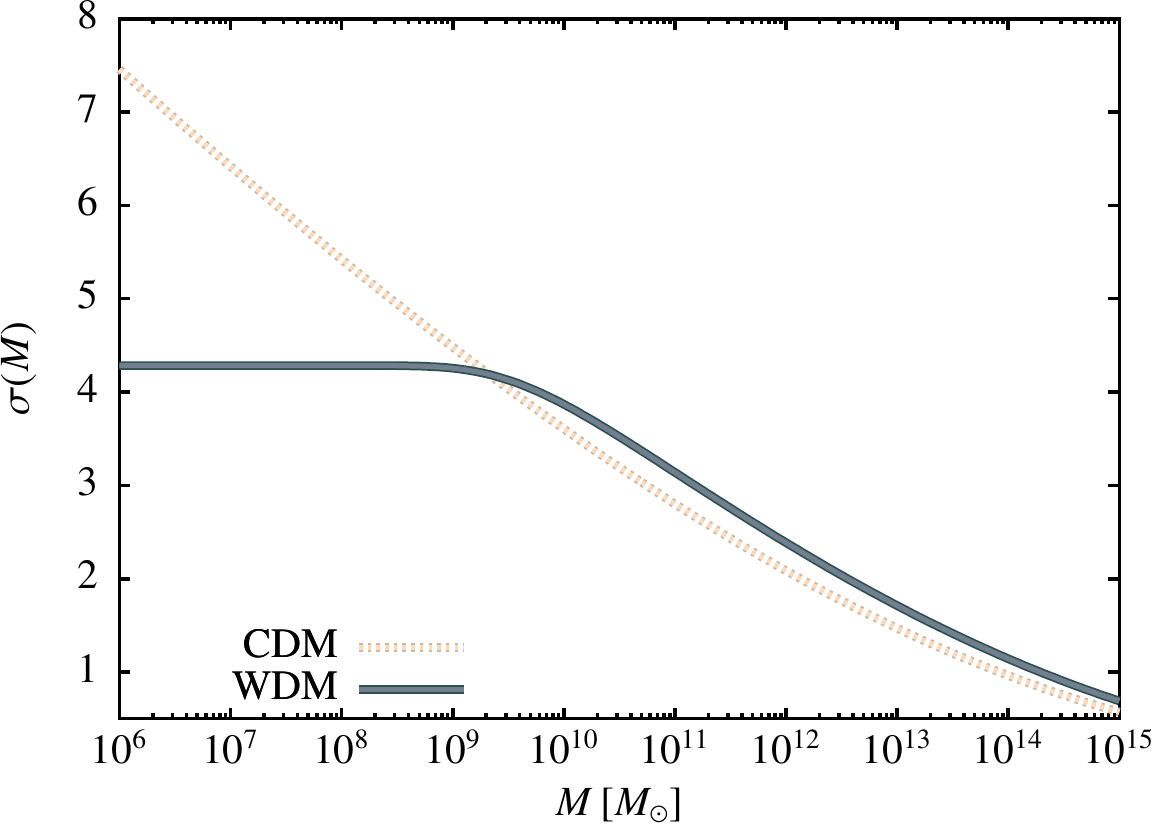}
\end{center}
\caption{The fractional root variance of the mass density field as a function of the mean mass contained within that filter. Both \protect\CDM\ and our canonical \protect\WDM\ cases are shown. A top-hat real-space window function is used for the \protect\CDM\ while a sharp $k$-space filter is used for \protect\WDM. The \protect\WDM\ case is suppressed below the \protect\CDM\ value for  $M\lesssim M_{\rm s}$.}
\label{fig:sigmaMassWdmVsCdm}
\end{figure}

Fig.~\ref{fig:sigmaMassWdmVsCdm} shows the resulting root variance in the density field (smoothed using a top-hat filter in real-space) as a function of mass enclosed within the filter for both \WDM\ and \CDM\ models. As expected, $\sigma(M)$ in our \WDM\ model is suppressed below the \CDM\ value for $M\lesssim M_{\rm s}$.

\subsection{Critical Overdensity for Collapse (Excursion Set Barrier)}\label{sec:barrier}

The Press-Schechter algorithm as formulated by \cite{bond_excursion_1991} associates the collapse of a dark matter halo with a random walk in $\delta(M)$ making its first crossing of some barrier, $B(S)$. This barrier is usually associated with the critical linear theory overdensity for collapse of spherical top-hat perturbations (although see \S\ref{sec:barrierRemapping} for a refinement of this assumption). In the \CDM\ case, that critical overdensity is independent of mass scale (since there are no preferred scales in the problem). In the non-\CDM\ case, this is no longer true. For example, with \WDM\ we expect collapse to become significantly more difficult on small scales, where the \WDM\ particles can stream out of collapsing overdensities due to their non-zero random velocities. \cite{barkana_constraints_2001} addressed the question of collapse thresholds in \WDM\ universes by performing a set of 1-D hydrodynamical simulations, in which pressure acted as a proxy for the velocity dispersion of \WDM\ particles. They find that the growth of collapsing overdensities is 
suppressed below a characteristic mass scale--i.e. the threshold for collapse increases rapidly with decreasing mass below the characteristics mass scale.

We find that the results of \cite{barkana_constraints_2001} can be well fit by the functional form\footnote{This fit is accurate for the regime where $\delta_{\rm c,WDM}/\delta_{\rm c,CDM} < 600$, but substantially over-predicts the results of \protect\cite{barkana_constraints_2001} for smaller masses. This is a deliberate choice--our aim was to match the shape of the function through the region where it transitions away from the \protect\CDM\ value. On smaller mass scales the suppression is so dramatic that the precise value of $\delta_{\rm c,WDM}/\delta_{\rm c,CDM}$ is unimportant.}:
\begin{eqnarray}
\delta_{\rm c,WDM}(M,t) &=& \delta_{\rm c,CDM}(t) \left\{ h(x) {0.04 \over \exp(2.3 x)} \right. \nonumber \\
& & \left. + [1-h(x)] \exp\left[{0.31687\over \exp(0.809 x)} \right] \right\}
\end{eqnarray}
where $x=\log(M/M_{\rm J})$, $M$ is the mass in question, $M_{\rm J}$ is the effective Jeans mass of the warm dark matter as defined by \citeauthor{barkana_constraints_2001}~(\citeyear{barkana_constraints_2001}; their eqn.~10):
\begin{eqnarray}
M_{\rm J} &=& 3.06 \times 10^8 \left( {1+z_{\rm eq} \over 3000}\right)^{1.5} \left({\Omega_{\rm M} h_0^2 \over 0.15}\right)^{1/2} \nonumber \\
&& \times \left({g_{\rm X} \over 1.5} \right)^{-1} \left({m_{\rm X}\over 1.0~\hbox{keV}}\right)^{-4}~{\rm M}_\odot,
\end{eqnarray}
the redshift of matter-radiation equality is given by
\begin{equation}
z_{\rm eq} = 3600 \left({\Omega_{\rm M} h_0^2 \over 0.15}\right)-1,
\end{equation}
and
\begin{equation}
 h(x) = 1/\{1+\exp[(x+2.4)/0.1]\}.
\end{equation}

\begin{figure}
\begin{center}
\includegraphics[width=8.5cm]{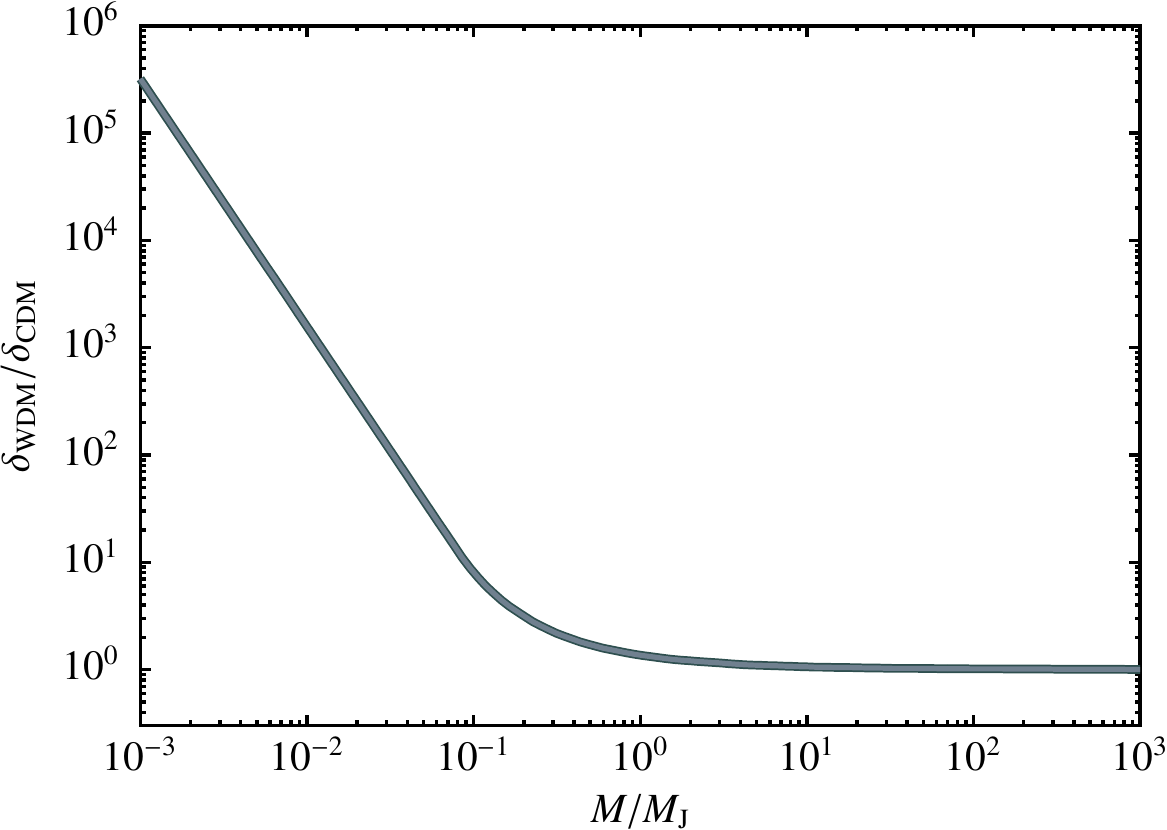}
\end{center}
\caption{The critical linear theory overdensity for the collapse of a spherical top-hat density perturbation in a \protect\WDM\ universe normalized to that in a \protect\CDM\ universe as a function of halo mass, $M$ (expressed in units of the effective Jeans mass of the \protect\WDM, $M_{\rm J}$). For $M/M_{\rm J}\gg 1$ the \protect\WDM\ and \protect\CDM\ cases coincide, but for $M/M_{\rm J}\lesssim 1$ \protect\WDM\ requires a much larger overdensity to undergo gravitational collapse.}
\label{fig:criticalOverdensity}
\end{figure}

The ratio of the resulting critical overdensity to the \CDM\ value with masses scaled to $M_{\rm J}$, is shown in Fig.~\ref{fig:criticalOverdensity}. For $M/M_{\rm J}\lesssim 1$ the critical overdensity for collapse in \WDM\ universes is much higher than in \CDM\ as a result of the non-zero velocity dispersion of warm dark matter particles. A small-scale perturbation must have a much larger density for its self-gravity to overcome this velocity dispersion and cause collapse.

We emphasize that the calculations of \cite{barkana_constraints_2001} are approximate as they are based on a hydrodynamical approximation which will not fully capture the collisionless dynamics of \WDM. We expect that the approximation made by \cite{barkana_constraints_2001} could lead to a small (order unity) difference in the characteristic mass scale for suppression of overdensity collapse (e.g. in a similar way that the Toomre stability threshold for collisionless stars and gas differ by a factor of $\pi/3.36$). There could plausibly be differences in detail in the shape of the collapse threshold as a function of mass, but we are unable to speculate what form those might take.  A more realistic calculation using a Boltzmann solver should be carried out to improve upon these results, and explore the dependence on cosmological parameters.

\subsubsection{Barrier Remapping}\label{sec:barrierRemapping}

In the \CDM\ case the original Press-Schechter algorithm as formulated by \cite{bond_excursion_1991} adopted a barrier, $B(S)$, for excursion sets equal to the critical linear-theory overdensity for the gravitational collapse spherical top-hat perturbations and which was independent of mass and equal to $\delta_{\rm c}(M[S],t) = \delta_{\rm c,0}/D(t)$ where $D(t)$ is the linear theory growth function\footnote{Note that, as is conventional, we place this time dependence into the critical overdensity, such that we can always work with $\sigma(M)$ at $z=0$. The growth function used here is the usual growth function computed for \protect\CDM, consistent with the definitions of \protect\cite{barkana_constraints_2001}.} and $\delta_{\rm c,0}$ is the collapse threshold (equal to $(3/20)(12\pi)^{2/3}\approx1.686$ in an Einstein-de Sitter universe; solutions for other cosmologies are given by \cite{percival_cosmological_2005} for example). It is now well known that this constant barrier does not result in a halo mass function that agrees well 
with that measured from N-body simulations of \CDM\ (e.g. \citealt{tinker_toward_2008}). Motivated by this discrepancy, \cite{sheth_ellipsoidal_2001} introduced 
a remapping of the barrier\footnote{This remapping was motivated by considerations of ellipsoidal collapse, the barrier for which differs from that for spherical collapse. In particular, \protect\cite{sheth_ellipsoidal_2001} noted that the density perturbations leading to low-mass halos are expected to be more ellipsoidal than those that give rise to the most massive halos--effectively making the collapse barrier a function of mass scale.} calibrated to improve the match:
\begin{equation}
B(S) \rightarrow B(S) \sqrt{A} \left(1 + b \left[S \over A B^2(S)\right]^c\right),
\end{equation}
where $A=0.707$, $b=0.5$ and $c=0.6$ are parameters that were tuned to obtain the best match. We retain this remapping of the barrier in this work\footnote{This mapping was derived strictly for the \protect\CDM\ case. We retain it here since for halos with masses, $M$, much greater than $M_{\rm s}$ we expect the \protect\WDM\ case to converge to the \protect\CDM\ solution. However, there is no guarantee that this mapping remains accurate for $M\lesssim M_{\rm s}$. In particular, it is possible that low mass halos close to the cut-off scale could be more spherical in \protect\WDM\ than in \protect\CDM\ due to the isotropizing effects of the \protect\WDM\ velocity dispersion. Calibration of our methods against reliable N-body simulations of \protect\WDM\ universes would be required to evaluate the accuracy of the remapping in this regime. The N-body simulations currently available and examined later in this work do not address this particular issue as they do not include the effects of the non-zero velocity dispersion of \protect\WDM. However, it is understood how to include velocity dispersions in \protect\WDM\ simulations \protect\citep{brandbyge_effect_2008}, and this should be included in future simulations.} when computing halo mass functions but \emph{not when computing halo merger rates} (see \S\ref{sec:mergerRates} for further discussion of this point).

\subsubsection{Barrier Scaling}\label{sec:barrierScaling}

As noted in \S\ref{sec:windowFunction} our choice of normalization for the window function used to derived $\sigma(M)$ from $P(k)$ results in $\sigma(M)$ for the \WDM\ case lying above the \CDM\ case for large masses. This will change the halo mass function for high mass halos--something which is not expected (i.e. \WDM\ should behave just like \CDM\ on sufficiently large scales). Therefore, we introduce a global re-scaling of the barrier (applied after the remapping described in \S\ref{sec:barrierRemapping}--an important point since that remapping is nonlinear), multiplying it by a factor of $1.197$. This value is chosen to counteract the higher $\sigma(M)$ on large scales\footnote{That is, the ratio of $\sigma(M)$ computed using our sharp $k$-space filter to that computed using the top-hat filter is $\sqrt{1.197}$. Note that this scaling factor does not constitute a free-parameter in our approach. Instead, it is fixed to offset the change in the $S(M)$ relation on large scales resulting from the difference between top-hat and sharp $k$-space window functions. Therefore, given a power spectrum and a value of $a$ (the coefficient in eqn.~\ref{eq:kToR}), the scaling factor is directly computable and uniquely determined.} and ensure that the mass function of high mass halos remains unchanged (see \S\ref{sec:windowFunction} for discussion of this point). This re-scaling of the barrier is always included, both when computing halo mass functions and also when computing halo merger rates. We note that this factor is very insensitive to cosmological parameters and $\sigma_8$ as it depends only upon the shape of the power spectrum (not the normalization). For example, if we compute the appropriate rescaling factor for WMAP-1 cosmological parameters (rather than the WMAP-7 cosmological parameters used throughout this work) we find that the factor changes by $<0.1\%$ (despite there being a  difference of 11\% in $\sigma_8$).

\begin{figure}
\begin{center}
\includegraphics[width=8.5cm]{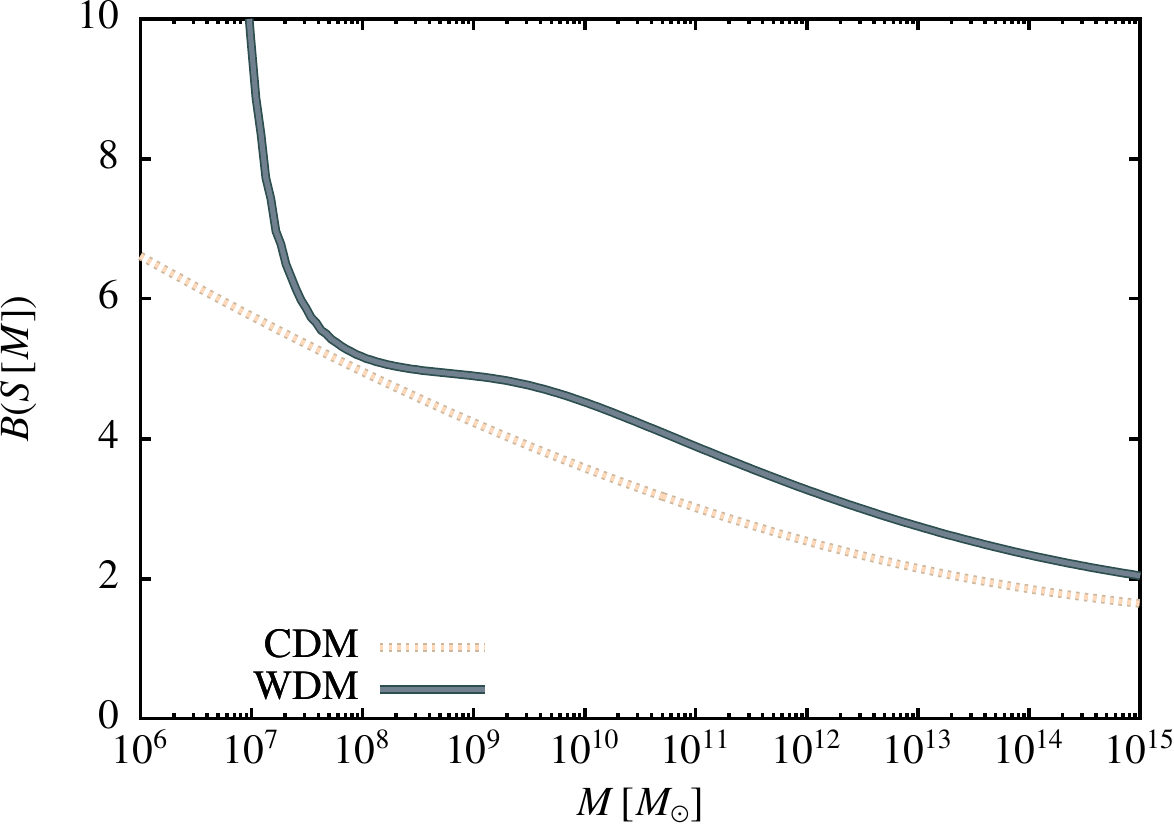}
\end{center}
\caption{The barrier for excursion sets in both the \protect\CDM\ and our canonical \protect\WDM\ cases. For large masses the two are offset by the constant factor of $1.197$ which accounts for the difference in definition (top-hat real space versus sharp $k$-space window functions) of $\sigma^2(M)$ in the two cases, but the \protect\WDM\ barrier becomes very large for $M\ll M_{\rm s}$. Below $M\sim M_{\rm s}$ the \protect\WDM\ barrier rises rapidly due to the velocity dispersion of \protect\WDM\ particles.}
\label{fig:barrierWdmVsCdm}
\end{figure}

The resulting excursion set barriers for both \CDM\ and our canonical \WDM\ model (including both remapping and subsequent re-scaling) are shown in Fig.~\ref{fig:barrierWdmVsCdm}. For large masses the two are offset by the factor of $1.197$, but the \WDM\ barrier becomes very large for $M\ll M_{\rm s}$ due to the velocity dispersion of the \WDM\ particles.

\subsection{Excursion Sets/Barrier Crossing}\label{sec:excursionSets}

Given a barrier, $B(S)$, and the variance of the density field, $S(M)$, the Press-Schechter algorithm proceeds by following random walks in $\delta(S)$ beginning from $(\delta,S)=(0,0)$. When a given random walk first crosses the barrier at some variance $S$, it is assumed that the point in question has collapsed into a halo of mass $M(S)$. The fraction of random walks crossing between $S$ and $S+\d S$, $f(S) \d S$, corresponds to the fraction of mass in the universe in halos of mass $M$ to $M+\d M$. 

We therefore must compute the probability for random walks to cross our barrier between $S$ and $S+\d S$. We will assume a Gaussian distribution of density perturbations, motivated by the simplest inflation models. In the case of a constant (or linear in $S$) barrier, there is a well-known analytic solution \citep{bond_excursion_1991,lacey_merger_1993,sheth_excursion_1998}. However, for an arbitrary barrier, the solution must be computed numerically. Our approach, described in detail in Appendix~\ref{app:numericalMethod}, is similar to that of \cite{zhang_random_2006} but is numerically more robust. Briefly, the excursion set problem requires finding a solution to an integral equation:
\begin{equation}
  1 =  \int_0^S f(S^\prime){\rm d}S^\prime + \int_{-\infty}^{B(S)} P(\delta,S) {\rm d} \delta,
 \label{eq:OldExcursionMethod}
\end{equation}
where $P(\delta,S) {\rm d} \delta$ is the probability for a trajectory to lie between $\delta$ and $\delta + {\rm d} \delta$ at variance $S$. This equation expresses mass conservation, i.e. at any $S$, all trajectories must either have crossed the barrier at some smaller $S$ (the first term in eqn.~\ref{eq:OldExcursionMethod}), or be below the barrier having never crossed at smaller $S$ (the second term in eqn.~\ref{eq:OldExcursionMethod}). To numerically solve this equation we discretize the variance, $S$, using a grid that is uniform in $S$ with a spacing of $\Delta S$. We then numerically solve eqn.~(\ref{eq:OldExcursionMethod}) as described in detailed in Appendix~\ref{app:numericalMethod}. The solution is extended to the maximum value of $S$, if such exists (as is the case in a \WDM\ model) or to sufficiently large $S$ that smaller mass scales are of no interest for the problem in question. The accuracy of our numerical solver is explored in Appendix~\ref{sec:firstCrossingNumerics}.

\subsection{Merger Rates and Progenitor Mass Functions}\label{sec:mergerRates}

The original Press-Schechter algorithm has been extended to compute conditional mass functions (i.e. the mass function of halos which will all belong to a single halo of larger mass at some later time; \citealt{lacey_merger_1993}). Further, the form of the conditional mass function in the limit of small timestep has been used to estimate merger rates of dark matter halos and so to construct merger trees \citep{cole_hierarchical_2000}. We therefore wish to compute the conditional mass function in the \WDM\ case, specifically in the limit of small $\delta t$. In the excursion set approach, the conditional mass function is found by simply solving the first crossing problem beginning from the $(\delta,S)$ of the parent halo and using a barrier corresponding to an earlier time. This is equivalent to solving the original first crossing problem with a modified barrier $B^\prime(S^\prime,t_1,t_0) = B(S^\prime+S,t_1)-B(S,t_0)$.

To compute merger rates we therefore solve the first crossing problem using our numerical method but with an effective barrier $B^\prime(S^\prime,t_1,t_0)$. We choose $t_1=(1-\epsilon)t_0$ where $\epsilon \ll 1$. The rate of crossing this effective barrier can then be estimated as $f(S)/\epsilon t_0$. We will explore the sensitivity of our results to the value of the numerical parameter, $\epsilon$. 

In this case the choice of grid in $S^\prime$ is particularly important. Accurate numerical solution requires many grid points at small $S^\prime$ (since $f(S^\prime)$ will peak very close to $S^\prime=0$ for small values of $\epsilon$), but also many points close to the maximum possible value of $S^\prime$. $S(M)$ becomes almost independent of $M$ close to the maximum value of $S$--as a result many points are required to resolve the cut-off of the halo mass function as a function of mass. Therefore, we adopt the following distribution for $S_i$:
\begin{eqnarray}
 S_0 &=& 0 \nonumber \\
 S_{i>0} &=& { 1 + 1/r \over 1/S_{i,{\rm lin}} + 1 / r S_{i,{\rm log}}},
\end{eqnarray}
where $S_{i,{\rm lin}}$ and $S_{i,{\rm log}}$ are sets of points spaced uniformly in $S$ and the logarithm of $S$ between the required minimum and maximum values and $r$ is a numerical parameter. For $r>1$ the resulting distribution of points is spaced uniformly in the logarithm of $S$ for small $S$, and transitions to being uniform in $S$ for large $S$--as a result, the grid has good resolution both close to $S=0$ and close to the maximum value of $S$. The value of $r$ controls the location of the transition between these two regimes. We find that a value of $r=10$ works well. The accuracy of our numerical solver for merger rates is explored in Appendix~\ref{sec:rateNumerics}.

When solving for halo merger rates we \emph{do not} include any remapping of the barrier function (as discussed in \S\ref{sec:barrierRemapping}). The remapping function was chosen to result in good agreement between the Press-Schechter halo mass function and that measured in the Millennium N-body simulations \citep{springel_simulations_2005}. However, our merger rates will be used in merger tree construction algorithms which, in the \CDM\ case, have been developed to work with a barrier with no remapping, but with their own empirical modification of merger rates designed to match progenitor mass functions found in N-body simulations (see \S\ref{sec:mergerTrees} for further discussion). Scaling of the barrier (see \S\ref{sec:barrierScaling}) is included however.

\subsection{Merger Tree Construction}\label{sec:mergerTrees}

To build merger trees, we follow the algorithm of \cite{parkinson_generating_2008}. Briefly, at each point in a merger tree, this algorithm evaluates the probability per unit time of a binary merger occurring along the branch
\begin{equation}
 {\d f \over \d \omega} = \int_{M_{\rm min}}^{M/2} {M \over M^\prime} {\d f \over \d t} {\d S \over \d M^\prime} \left| {\d t \over \d \omega}\right| G[\omega,\sigma(M),\sigma(M^\prime)] \d M^\prime,
\end{equation}
where $\omega = \delta_{\rm c,0}/D(t)$ and $\delta_{\rm c,0}$ is the critical overdensity for collapse in the \CDM\ case and the integration is from the lowest mass halo to be resolved in the tree, $M_{\rm min}$, to halos of half the mass of the current halo, corresponding to an equal mass merger. \cite{cole_hierarchical_2000} discuss the subtleties of why the integration is carried out over the lower half of the progenitor mass range. The rate of accretion of mass in halos below the resolution limit of the merger tree
\begin{equation}
 {\d R \over \d \omega} = \int_0^{M_{\rm min}} {\d f \over \d t} {\d S \over \d M^\prime} \left| {\d t \over \d \omega}\right| G[\omega,\sigma(M),\sigma(M^\prime)] \d M^\prime,
 \label{eq:unresolvedAccretion}
\end{equation}
where in this case the integral is taken over all unresolved halos (i.e. those less massive than $M_{\rm min}$). In the above, $G[\omega,\sigma(M),\sigma(M^\prime)]$ is an empirical modification of the progenitor mass function introduced to obtain results consistent with those from N-body simulations of \CDM\ universes. \cite{parkinson_generating_2008} show that the form:
\begin{equation}
 G[\omega,\sigma_1,\sigma_2] = G_0 \left({\sigma_2 \over \sigma_1}\right)^{\gamma_1}  \left({\omega \over \sigma_1}\right)^{\gamma_2},
\end{equation}
with $G_0=0.57$, $\gamma_1=0.38$ and $\gamma_2=-0.01$ provides a good match to progenitor mass functions measured from the Millennium Simulation \citep{springel_simulations_2005}, which assumes similar cosmological parameter values as are used in this work. The validity of this empirical modification on much smaller mass scales than are probed by the Millennium Simulation is explored in Appendix~\ref{sec:pchNumerics}.

We retain this same empirical modification in this work since it should remain valid for \WDM\ in the limit of high-mass halos. For masses comparable to and below $M_{\rm s}$ this empirical modification may no longer be accurate. Our justification for retaining this empirical modification is the good agreement achieved with progenitor mass functions from N-body simulations of \WDM\ as will be shown in \S\ref{sec:Nbody}.

A timestep is then chosen that is sufficiently small such that the probability of multiple branchings is small and the fractional change in mass due to subresolution accretion is small. This timestep is then taken, branching if a random deviate lies below the branching probability, and with mass removed at the rate of subresolution accretion. If branching does occur the mass of one of the branched halos is selected at random from the distribution $\d^2 f/\d\omega \d M^\prime$ and the other is chosen to ensure mass conservation.

\subsubsection{Smooth Accretion}\label{sec:smoothAccretion}

In the \CDM\ case, all mass is locked into halos on some scale\footnote{Assuming that $\sigma(M)$ continues to rise monotonically to arbitrarily small scales. In practice this is not true as even \protect\CDM\ will have some cut-off in its power spectrum on very small scales \protect\citep{green_power_2004,loeb_small-scale_2005}. However, for our present purposes the assumption that all mass is locked into halos in \protect\CDM\ is a good one.}. This implies that the progenitor mass function in \CDM\ contains the entire mass of the parent halo, i.e.:
\begin{equation}
\int_0^\infty f(S^\prime) \d S^\prime = 1.
\end{equation}
For the \WDM\ case this is not true and we find,
\begin{equation}
\int_0^{S_{\rm max}} f(S^\prime) \d S^\prime < 1,
\end{equation}
i.e. not all trajectories cross the barrier. We identify these trajectories as corresponding to smooth accretion, i.e. accretion of mass which does not belong to any collapsed halo. This is physically distinct from the unresolved halos accounted for by eqn.~(\ref{eq:unresolvedAccretion}), but we can nevertheless account for this smooth accretion by boosting $\d R / \d \omega$ by an amount
\begin{equation}
 \left. {\d R \over \d \omega}\right|_{\rm smooth} = {\d f_{\rm n} \over \d t} \left| {\d t \over \d \omega}\right| G[\omega,\sigma(M),\sqrt{S_{\rm max}}],
\end{equation}
where $f_{\rm n}=1-\int_0^{S_{\rm max}} f(S^\prime) \d S^\prime$. Note that we choose to include the empirical correction of \cite{parkinson_generating_2008} here, using the maximum value of $\sigma(M)$ for the $\sigma_{\rm C}$ argument. This ensures a treatment consistent with resolved halos but, once again, should be tested and calibrated against \WDM\ simulations. The importance of this smooth accretion is explored in Appendix~\ref{sec:treeNumerics}.

\section{Results for Warm Dark Matter}\label{sec:wdmResults}

The methods described in \S\ref{sec:methods} have been implemented within the open source semi-analytic galaxy formation code, {\sc galacticus} \citep{benson_galacticus:_2012}. All results presented in this section were generated using {\sc galacticus} {\tt v0.9.1 r903}. Control files and scripts to generate all results presented in this paper using {\sc galacticus} can be found at \href{http://users.obs.carnegiescience.edu/abenson/galacticus/parameters/dmMergingBeyondCDM1.tar.bz2}{{\tt http://users.obs.carnegiescience.edu/abenson/\newline galacticus/parameters/dmMergingBeyondCDM1.tar.bz2}}.

\begin{figure}
\begin{center}
\includegraphics[width=8.5cm]{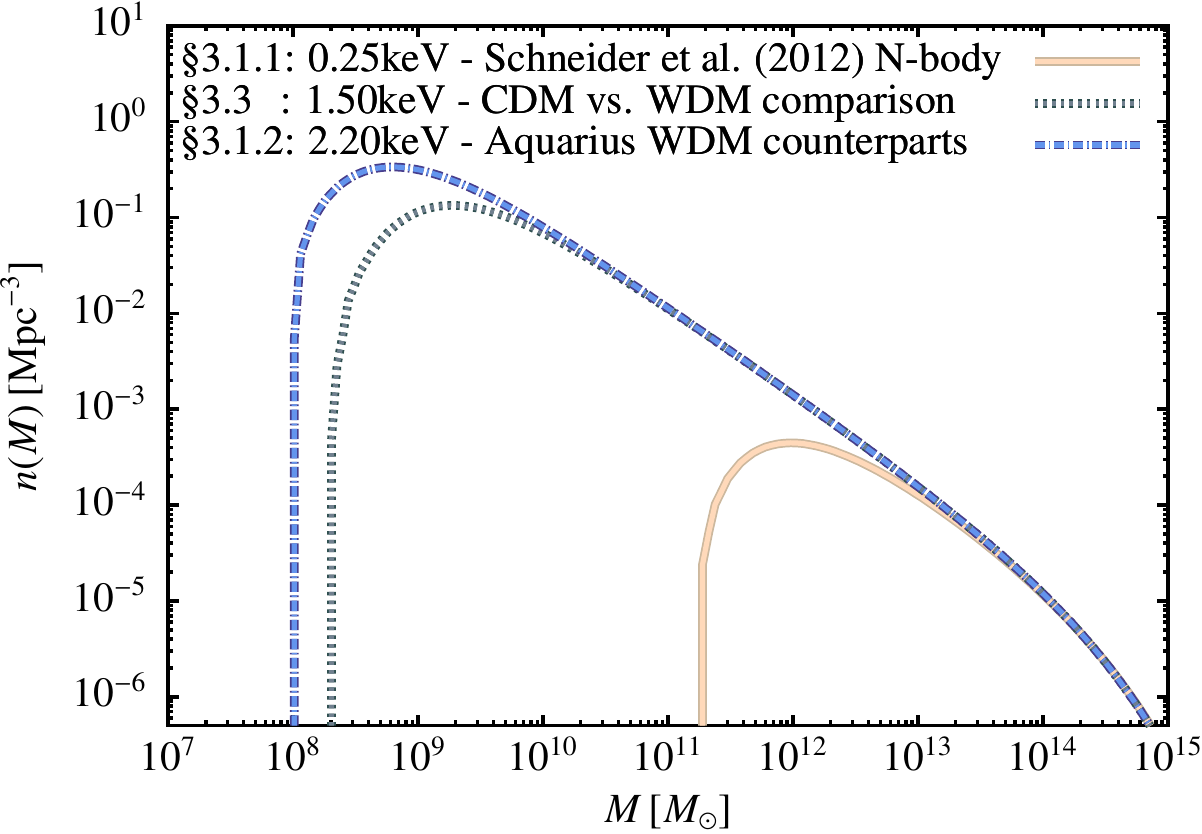}
\end{center}
\caption{Dark matter halo mass functions, computed using the methods described in \S\protect\ref{sec:methods}, for the three different \protect\WDM\ particle masses considered in this work. The labels give the particle mass and the section of this work in which that mass is used.}
\label{fig:massFunctionCompare}
\end{figure}

We consider three different \WDM\ particle masses. In \S\ref{sec:NbodyMassFunction} we use $m_{\rm X}=0.25$~keV to match the N-body simulations of \cite{schneider_non-linear_2012}, in \S\ref{sec:NbodyProgenitor} we use $m_{\rm X}=2.2$~keV to match the Aquarius \WDM\ counterpart simulations (described below), and in \S\ref{sec:wdmVsCdm} we use $m_{\rm X}=1.5$~keV in comparisons of \WDM\ and \CDM. For reference, Fig.~\ref{fig:massFunctionCompare} shows the halo mass function in these three cases, illustrating the expected increase in the cut-off mass, $m_{\rm s}$, as the \WDM\ particle mass is decreased.

\subsection{Comparison with N-body Simulations}\label{sec:Nbody}

\subsubsection{Halo Mass Function}\label{sec:NbodyMassFunction}

\begin{figure}
\begin{center}
\includegraphics[width=8.5cm]{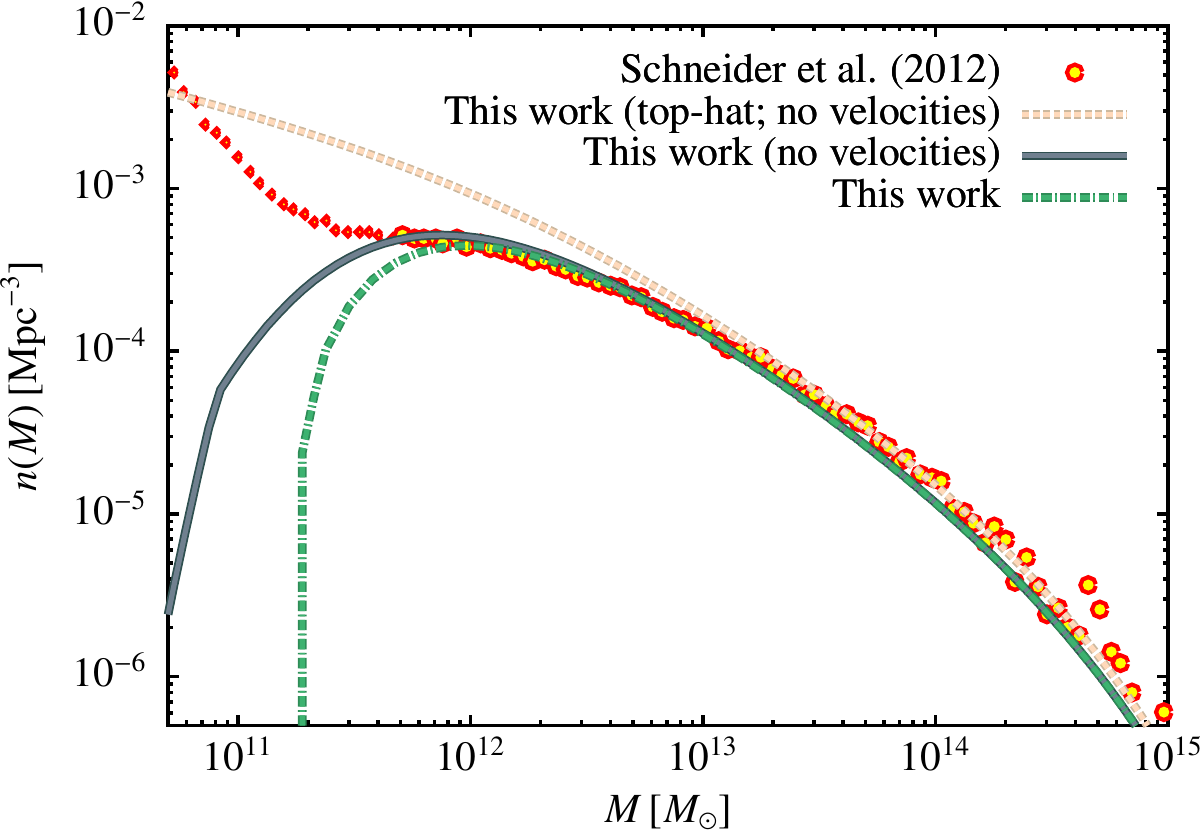}
\end{center}
\caption{The dark matter halo mass function for \protect\WDM\ in the case of a $0.25$~keV particle. Points show the results of an N-body simulation of \protect\WDM\ from \protect\cite{schneider_non-linear_2012} (small points are used to indicate the region in which the N-body simulation results are unreliable as a result of being dominated by halos formed through artificial fragmentation), while the line shows the result from this work with dark matter properties and cosmological parameters matched to those used by \protect\cite{schneider_non-linear_2012}. Note that the upturn in the N-body mass function below $2\times10^{11}$~M$_\odot$ arises to artificial fragmentation of filaments (see \protect\citealt{wang_discreteness_2007}).}
\label{fig:schneiderMassFunction}
\end{figure}

N-body simulations should, in principle, provide an accurate determination of the dark matter halo mass function in \WDM\ cosmologies, provided that initial conditions are constructed carefully. Points in Fig.~\ref{fig:schneiderMassFunction} show the mass function measured in an N-body simulation of $0.25$~keV \WDM\ carried out by \cite{schneider_non-linear_2012}. The upturn below $2\times 10^{11}$~M$_\odot$ is a numerical artifact, arising from the fragmentation of filaments due to particle discreteness \citep{wang_discreteness_2007}. This is a challenging problem for N-body simulations of \WDM\ as the mass scale at which the upturn appears decreases only as $N^{-1/3}$ (where $N$ is the particle number).

The solid lines in Fig.~\ref{fig:schneiderMassFunction} shows the mass function predicted by our calculation using the same cosmological and \WDM\ particle parameters as \cite{schneider_non-linear_2012}. \cite{schneider_non-linear_2012} identified halos using a friends-of-friends algorithm with a linking length of $b=0.2$, corresponding approximately to density contrasts of $200$. In our model, the relevant density contrasts are those arising from the spherical top hat collapse model (e.g. \citealt{percival_cosmological_2005}). We correct the masses reported by \cite{schneider_non-linear_2012} for this difference by assuming the halos have NFW density profiles \citep{navarro_universal_1997} with concentrations given by the \CDM\ results of \cite{gao_redshift_2008} modified by the \WDM-to-\CDM\ conversion factor reported by \cite{schneider_non-linear_2012}. Additionally, \cite{schneider_non-linear_2012} do not include any velocity dispersion of dark matter particles in their initial conditions and so the fairest comparison is one in which we do not modify the critical overdensity for collapse as described in \S\ref{sec:barrier}.

Results of three different calculations from this work are shown in Fig.~\ref{fig:schneiderMassFunction}. Going from top to bottom, the first two lines do not include the effects of velocity dispersion (consistent with \citealt{schneider_non-linear_2012}). The first line corresponds to a case where we use a top-hat real-space window function to determine $\sigma(M)$. It can be seen that this curve, while in good agreement with the N-body results for high masses, does not produce sufficient suppression of the mass function at lower masses--a similar discrepancy was seen by \cite{barkana_constraints_2001} when comparing their Press-Schechter-based model for \WDM\ halo formation to the N-body simulations of \cite{bode_halo_2001}. The second line (blue) switches to using a sharp $k$-space window function as described in \S\ref{sec:windowFunction}. For high mass halos, this results in only a small change in the mass function compared to the \CDM\ case due to the difference in $S(M)$ when computed with sharp $k$-space and top-hat window functions (even after scaling the barrier height to compensate for this difference as much as possible; see \S\ref{sec:barrierScaling})\footnote{The slight difference between this curve and that computed using a top-hat window function is due our inclusion of the \protect\cite{sheth_ellipsoidal_2001} remapping. If this were not included, the factor of $1.197$ increase the barrier would precisely compensate for the difference in $\sigma(M)$ between these two curves on large scales. With the \protect\cite{sheth_ellipsoidal_2001} remapping included this constant factor cannot correct for the offset fully (due to the non-linear nature of the \protect\cite{sheth_ellipsoidal_2001} remapping). Future, high-precision work should consider re-calibrating the \protect\cite{sheth_ellipsoidal_2001} remapping to match a \protect\CDM\ halo mass function with $\sigma(M)$ computed using our sharp $k$-space window function. This would obviate the need for a separate multiplicative increase in the barrier and could better capture the scale-dependence of the required correction.}, but results in an excellent match to the suppression of the abundance of low mass halos, indicating that the discrepancies found in previous works were due to the artificial increasing in $\sigma(M)$ below the cut-off which arises when a top-hat real-space filter is used. At the lowest masses the suppression of the mass function is masked in the N-body simulation as it is masked by the upturn due to artificial fragmentation of filaments. Finally, the green line adds in the effects of velocity dispersion (which are not included in the N-body simulation), illustrating the importance of this effect to accurately model the suppression of the lowest mass halos.

\subsubsection{Progenitor Mass Functions/Merger Rates}\label{sec:NbodyProgenitor}

\begin{figure*}
\begin{center}
\begin{tabular}{cc}
\includegraphics[width=8.5cm]{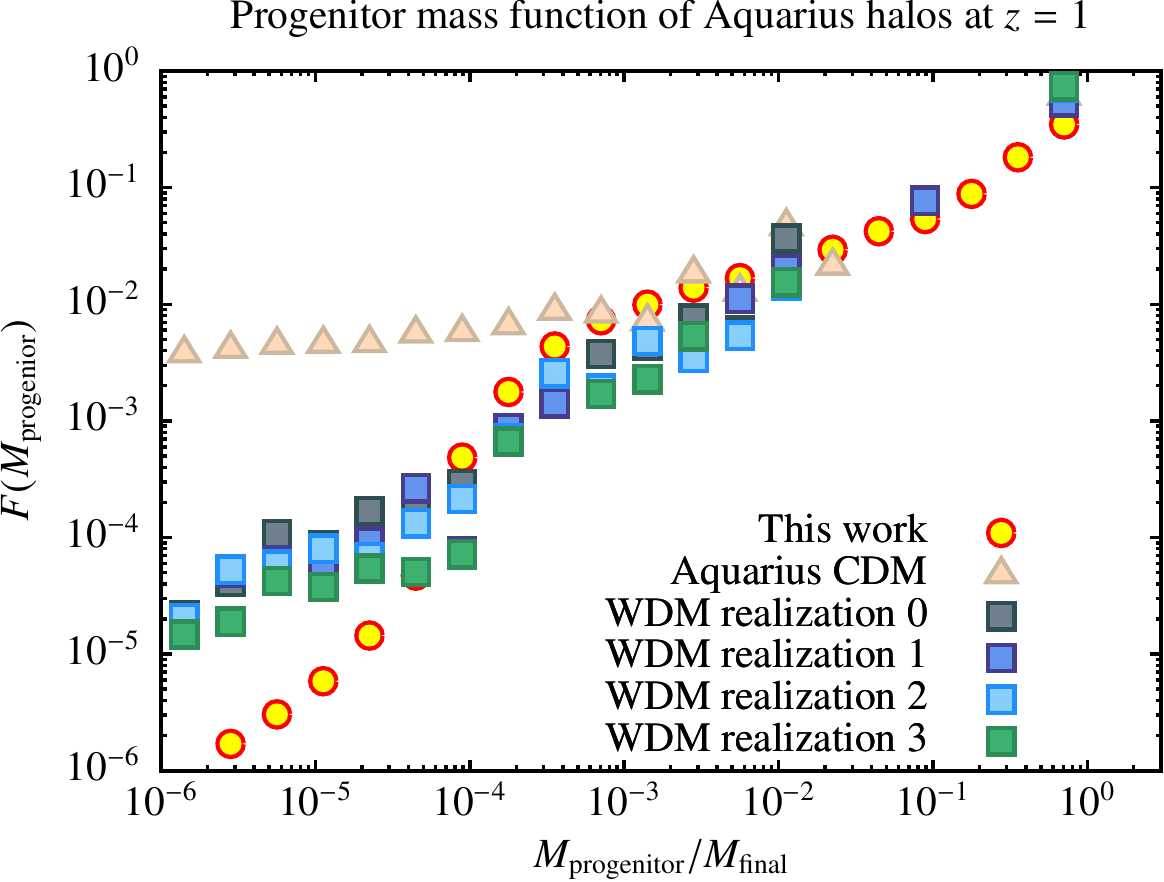} &
\includegraphics[width=8.5cm]{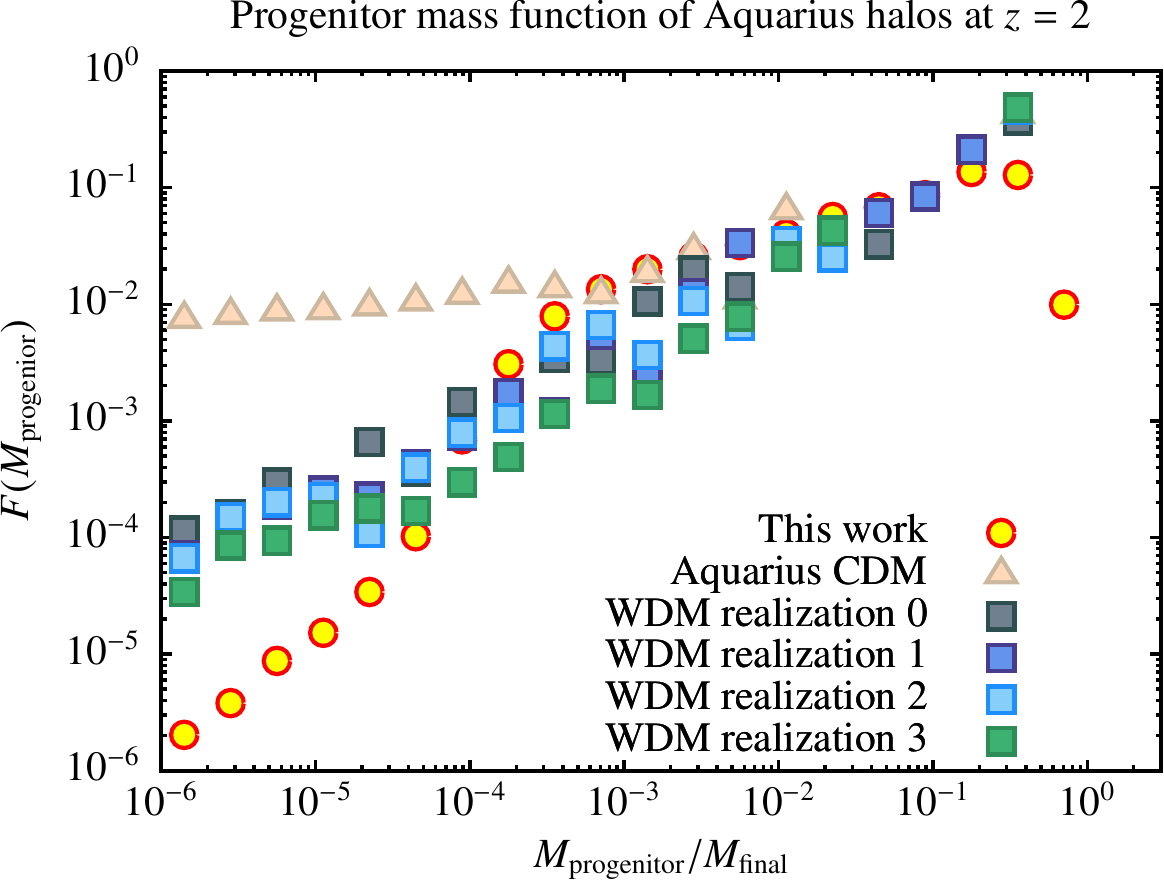}
\end{tabular}
\end{center}
\caption{Progenitor mass functions of four realizations of the Milky Way-mass \protect\WDM\ dark matter halos at $z=1$ and 2 (squares) compared with the predictions of this work (circles). For reference, the corresponding progenitor mass functions from one realization of the Aquarius \protect\CDM\ simulations are shown as triangles. Each panel shows the fraction of the halo's mass contributed by progenitors in each mass bin. Masses are shown as a fraction of the final halo mass.}
\label{fig:aquariusWDM}
\end{figure*}

A set of \WDM\ Milky Way mass haloes have been simulated and analyzed in Lovell et al (in prep). The haloes they simulated are \WDM\ counterparts of the Aquarius \CDM\ haloes presented in \cite{springel_aquarius_2008}, and so have masses of order $10^{12}M_\odot$--we will therefore refer to them as ``Milky Way-mass \WDM\ simulations''. Here we make use of a set of four (A-D) haloes simulated with approximately 40~million particles within their virial radii (resolution level 3 in the notation of the Aquarius project). One of the haloes (A) has also been run at higher (level 2) resolution and we have used this to verify that the progenitor mass functions that we show in Fig.~\ref{fig:aquariusWDM} have accurately converged over the range of masses plotted. The cosmological parameters for this set of simulations were chosen to match the WMAP7 results of \cite{komatsu_seven-year_2011}, however the \CDM\ power spectrum was modified using the prescription of \cite{bode_halo_2001} for \WDM\ as given in our eq. (\ref{eq:transferFunction}) with a smoothing scale of $\lambda_{\rm s}=78.4$~kpc, corresponding to an approximately $2.2$~keV thermal \WDM\ particle. In each of the simulations haloes and subhaloes were identified at each of the $128$ output times using the FOF \citep{davis_evolution_1985} and SUBFIND \citep{springel_populating_2001} algorithms respectively. Merger trees of subhaloes were constructed by identifying the descendant of each subhalo and halo merger trees built from the subhalo membership of each halo as described in Jiang et al. (in prep) and \cite{merson_lightcone_2012}.  These halo merger trees can be
thought of as FOF halo merger trees that have been cleaned up to avoid
problems that occur when FOF haloes are essentially composed of two
distinct haloes linked by a low density bridge.

The issue of spurious
haloes formed by the fragmentation of filaments due to numerical noise
\citep{wang_discreteness_2007} is investigated in Lovell et al. (in prep). They find that most of such (sub)haloes can be identified by looking at the shape in the initial conditions of the region from which their particles originated. The spurious (sub)haloes typically originate from very flattened configurations. Lovell et al. determine a
threshold on the axis ratio, $c/a$, of the inertia tensor of the
initial particle distribution such that for a \CDM\ simulation 99\% of haloes pass this cut while in a \WDM\ simulation most of the spurious haloes are excluded. Here we use this criterion to exclude complete subhalo branches from the N-body merger trees. In these N-body merger trees the same subhalo is identified at subsequent epochs by tracing the fate of a fraction of its most bound particles. In this way we can identify the point at which a halo dissolves as a result of disruption within a larger (sub)halo and form a branch consisting of its main progenitor at all earlier times. We discard such a branch if at the point at which this subhalo had half its maximum mass the subhalo fails the cut on initial axis ratio.

Fig.~\ref{fig:aquariusWDM} compares the progenitor mass functions measured from a large number of merger tree realizations generated using the techniques described in this work (circles) compared with progenitor mass functions measured from four Milky Way-mass \WDM\ simulation halos (squares). For reference, the equivalent \CDM\ progenitor mass function from a single realization of the Aquarius simulations is shown (triangles). Given the halo-to-halo scatter, there is good agreement down to $M_{\rm progenitor}/M_{\rm final}\approx 10^{-4}$. Below this, the abundance of progenitors in the \WDM\ N-body simulations exceeds that predicted by our techniques. At these mass scales, the artificial halo rejection algorithm reduces the number of progenitors by over an order of magnitude. This excess of low mass progenitor halos is consistent with failure rate for the artificial halo rejection algorithm of slightly less than 10\%. We cannot rule out such a failure rate in the algorithm and so the true progenitor mass function could decline much more rapidly. The progenitor mass functions from our analytic calculations decline rapidly as they approach $M_{\rm progenitor}/M_{\rm final}\approx 10^{-4}$, but then transition to a slower, power-law decline at lower masses. As will be discussed in \S\ref{sec:mergerTreeStats}, this appears to be due to the finite difference approximation used to compute merger rates (see \S\ref{sec:mergerRates}) and so can be suppressed as necessary by lowering the value of $\epsilon$.

In the \CDM\ case the mass function (expressed in this way) levels off to be almost constant below $M_{\rm progenitor}/M_{\rm final}\sim 10^{-3}$. In our calculation, which tracks the \CDM\ case extremely well at high masses, this flattening begins (for $M_{\rm progenitor}/M_{\rm final}\gtrsim 10^{-3}$) but is then interrupted by the cut-off due to \WDM\ physics. The result is a ``bump'' feature. The presence of such a bump is less clear in the \WDM\ N-body progenitor mass functions (although there is a hint of it in the $z=1$ results), due to the noisiness of those results. This feature may have interesting implications for the expected number of surviving dwarf scale subhalos, and so represents an interesting avenue for future investigation.

\subsection{Limitations of Only Modifying Power Spectrum}\label{sec:limits}

\begin{figure}
\begin{center}
\includegraphics[width=8.5cm]{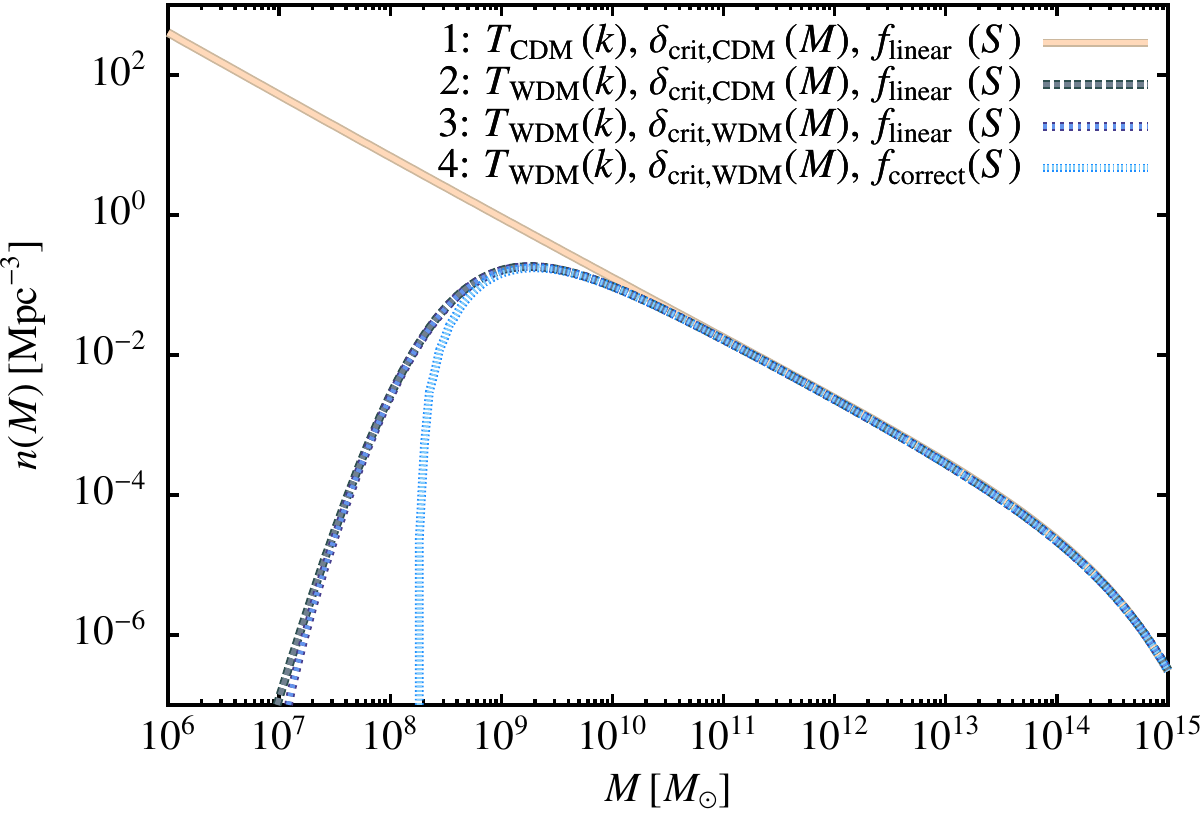}
\end{center}
\caption{The dark matter halo mass function for \protect\CDM\ (line 1) and for various approximations to the $1.5$~keV \protect\WDM\ solution. Line 2 shows the result of using the \protect\WDM\ transfer function, but retaining the \protect\CDM\ mass-independent critical overdensity and the functional form of $f(S)$ appropriate to a linear barrier, $f_{\rm linear}(S)$. Line 3 improves on this approximation by using the correct mass-dependent \protect\WDM\ critical overdensity, but still uses $f_{\rm linear}(S)$. Finally, line 4 shows the result of using the correct functional form of $f(S)$ for the \protect\WDM\ barrier. Note that, for large halo masses, all lines coincide precisely and so are hidden beneath Line 4.}
\label{fig:warmDarkMatterWrongSolutions}
\end{figure}

Having demonstrated that our model is consistent with the available N-body simulation we now consider the effects of treating \WDM\ incorrectly or incompletely in the extended Press-Schechter approach, as has previously happened in the literature \citep{menci_galaxy_2012}. Fig.~\ref{fig:warmDarkMatterWrongSolutions} shows a series of results for the halo mass function for $1.5$~keV \WDM. For reference, line 1 shows the mass function for \CDM\ derived assuming a mass-independent critical overdensity. Line 2 switches to using the \WDM\ power spectrum, but retains the \CDM\ mass-independent critical overdensity. There is a clear suppression of the mass function below $M_{\rm s}$. Line 3 now uses the \WDM\ power spectrum and adds mass dependence to the critical overdensity, but still uses the barrier crossing solution for a linear barrier\footnote{Specifically, we adopt a $B(S)$ corresponding to the mass-dependent \protect\WDM\ critical overdensity, but simply use it in the solution for $f(S)$ appropriate to a linear 
barrier.} given in eqn.~(\ref{eq:LinearBarrierSolution}). The suppression of the mass function is almost unchanged compared to the previous case where we used the \CDM\ critical overdensity. Finally, line 4 shows the result of switching to using the correct, numerically determined, functional form for $f(S)$. The mass scale of the cut-off undergoes a dramatic shift of almost an order of magnitude to higher mass compared to the previous case. The reason for this is simple to understand. When using the \WDM\ $B(S)$ in $f_{\rm linear}(S)$ we are implicitly assuming that the barrier was flat at the same value of $B(S)$ at all smaller $S$ (this being the assumption used to derive $f_{\rm linear}(S)$. This gives a certain crossing probability. When we numerically determine $f_{\rm correct}(S)$ the full $S$-dependence of the barrier is taken into account. In the first case, a random walk crossing $B(S)$ at $S$ must have remained below $B(S)$ for all $S^\prime < S$, while in the second case it must have remained below $B(S^\prime)$ for all $S^\prime < S$. Since $B(S^\prime) < B(S)$ for all $S^\prime < S$ this 
second 
condition is much more stringent, and so far fewer random walks will satisfy it. Thus, the crossing probability, and so the mass function, will be more strongly suppressed. This illustrates the importance of correctly solving the barrier crossing problem for \WDM\ calculations.

\subsection{Warm vs. Cold Dark Matter}\label{sec:wdmVsCdm}

In the following we compare example results for \CDM\ and $1.5$~keV \WDM\ cases. This \WDM\ particle mass differs from those used in previous sections (where the choice was constrained to match those assumed in different N-body simulations). In all cases, we use the correct $f(S)$ (determined numerically for the \WDM\ case) and include remapping of the barrier for calculations of first crossing probabilities and mass functions.

\subsubsection{Mass Functions}

\begin{figure}
\begin{center}
\includegraphics[width=8.5cm]{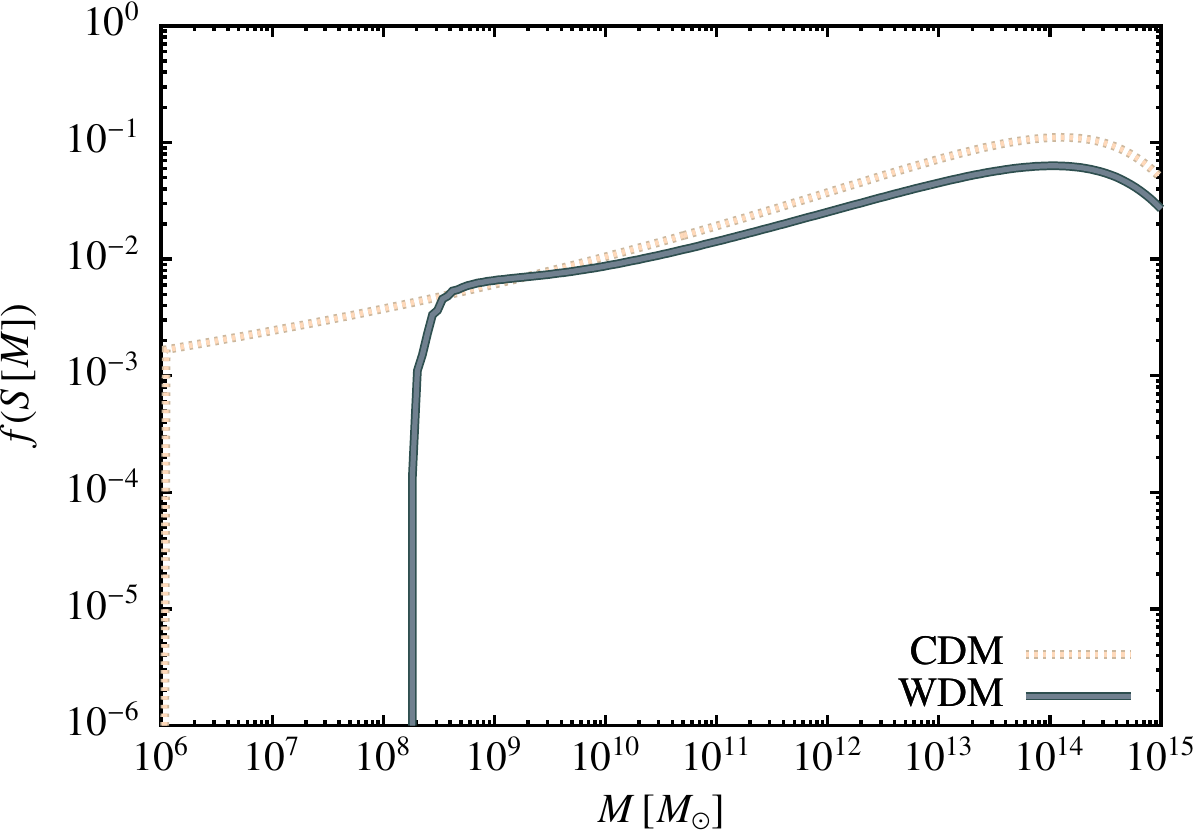}
\end{center}
\caption{The first crossing probability, $f(S)$, is shown as a function of mass, $M$, for both \protect\CDM\ and \protect\WDM\ cases for $z=0$.}
\label{fig:firstCrossingWdmVsCdm}
\end{figure}

Fig.~\ref{fig:firstCrossingWdmVsCdm} shows the first crossing probability, $f(S)$, for both \CDM\ and \WDM\ as a function of mass, $M$. The two are offset by a constant multiplicative factor at large masses (small $S$). This is expected--we have chosen to re-scale the \WDM\ barrier such that the \emph{halo mass function} remains unchanged relative to the \CDM\ expectation for large masses when we use the sharp $k$-space filter in our \WDM\ calculations. The mapping from $f(S)$ to the mass function, $n(M)$, is proportional to ${\rm d}S/{\rm d} M$:
\begin{equation}
 n(M) = {\bar{\rho} \over M} f(S[M]) {{\rm d} S\over {\rm d} M}.
\end{equation}
For large masses, ${\rm d}S/{\rm d} M$ is larger at fixed mass for large masses in our \WDM\ calculations compared to the equivalent \CDM\ calculation (a consequence of the different window functions adopted for the two cases). This difference in ${\rm d}S/{\rm d} M$ offsets the difference in $f(S)$ for large masses resulting in halo mass functions that agree between \WDM\ and \CDM.

Below around $M_{\rm s}$ the \WDM\ first crossing probability is suppressed due to the rapidly rising barrier $B(S)$. (There is a small region where the \WDM\ $f(S)$ exceeds that of \CDM\ due to differences in the mapping from $M$ to $S$ in the two cases.

These first crossing distributions translate directly to halo mass functions, as shown by Lines 1 and 4 of Fig.~\ref{fig:warmDarkMatterWrongSolutions}. As expected, the suppression in $f(S)$ in \WDM\ translates to a strong suppression in the mass function below around $M_{\rm s}$. At larger masses, the two are indistinguishable.

\subsubsection{Merger Rates}

\begin{figure}
\begin{center}
\includegraphics[width=8.5cm]{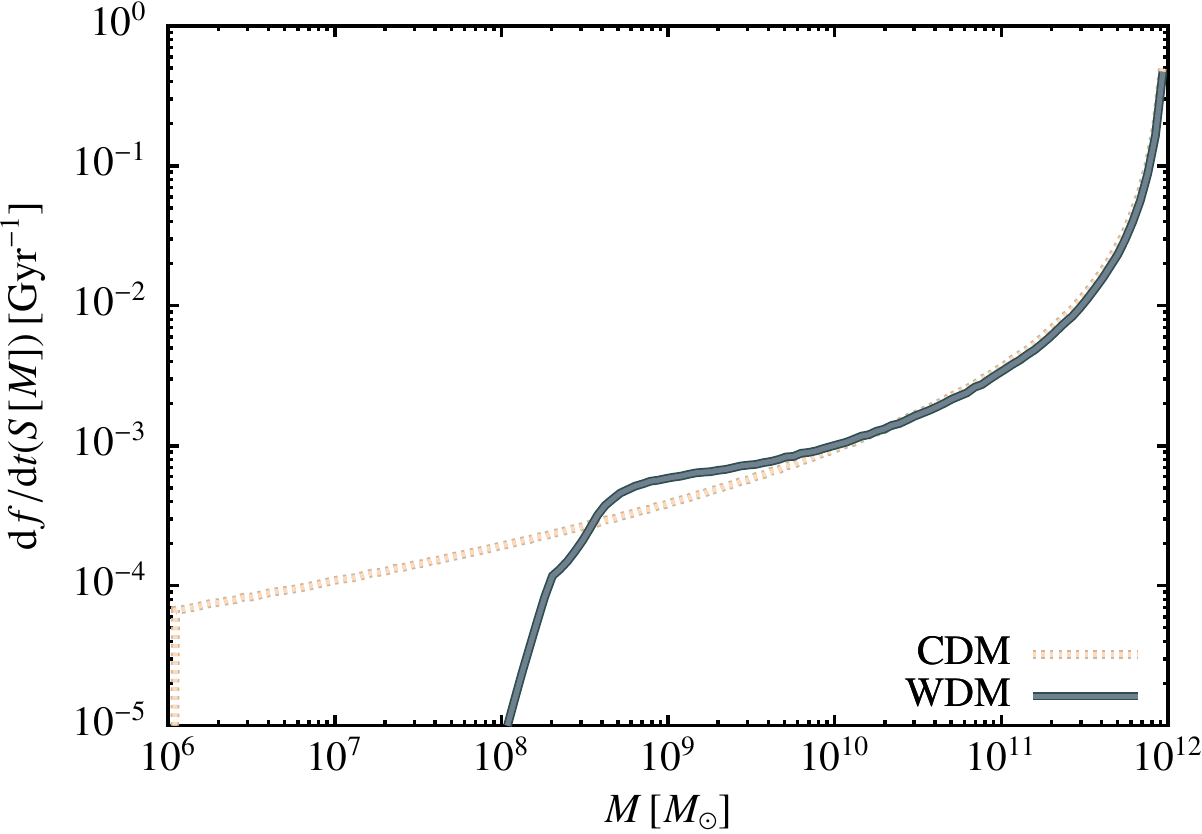}
\end{center}
\caption{The rate of first crossing, $\d f/\d t (S)$, is shown as a function of mass, $M$, for both \protect\CDM\ and \protect\WDM\ cases, for the conditional barrier appropriate to a $10^{12}$~M$_\odot$ halo at $z=0$.}
\label{fig:firstCrossingRatesWdmVsCdm}
\end{figure}

Fig.~\ref{fig:firstCrossingRatesWdmVsCdm} shows the rate of first crossing for \CDM\ and \WDM\ barriers for a $10^{12}$~M$_\odot$ halo at $z=0$. The two lines almost coincide at high mass, although the \WDM\ line lies slightly below the \CDM\ line\footnote{For the same reason as in Fig.~\protect\ref{fig:firstCrossingWdmVsCdm}. Here the difference is less evident, partly because the two lines are closer to vertical, but also because the largest mass scale showed here is lower that in Fig.~\protect\ref{fig:firstCrossingWdmVsCdm} such that $S(M)$ differs less from \protect\WDM\ to \protect\CDM\ cases.}. The merger rate in the \WDM\ case is strongly suppressed below around $M_{\rm s}$ due to the lack of halo in that mass range. The slight enhancement in the rate of first crossing in \WDM\ compared to \CDM\ just above $M_{\rm s}$ is due to the different mapping between mass and variance in the two cases.

\subsubsection{Merger Tree Statistics}\label{sec:mergerTreeStats}

Using merger rates computed as described in \S\ref{sec:mergerRates} we construct merger trees in both \CDM\ and \WDM\ cases using the algorithm described in \S\ref{sec:mergerTrees}, beginning with a halo of mass $10^{12}$~M$_\odot$ at $z=0$. We generate $1743$ trees in each case and construct the mean progenitor mass function and the mean mass accretion history.

\begin{figure}
\begin{center}
\begin{tabular}{c}
\includegraphics[width=8.5cm]{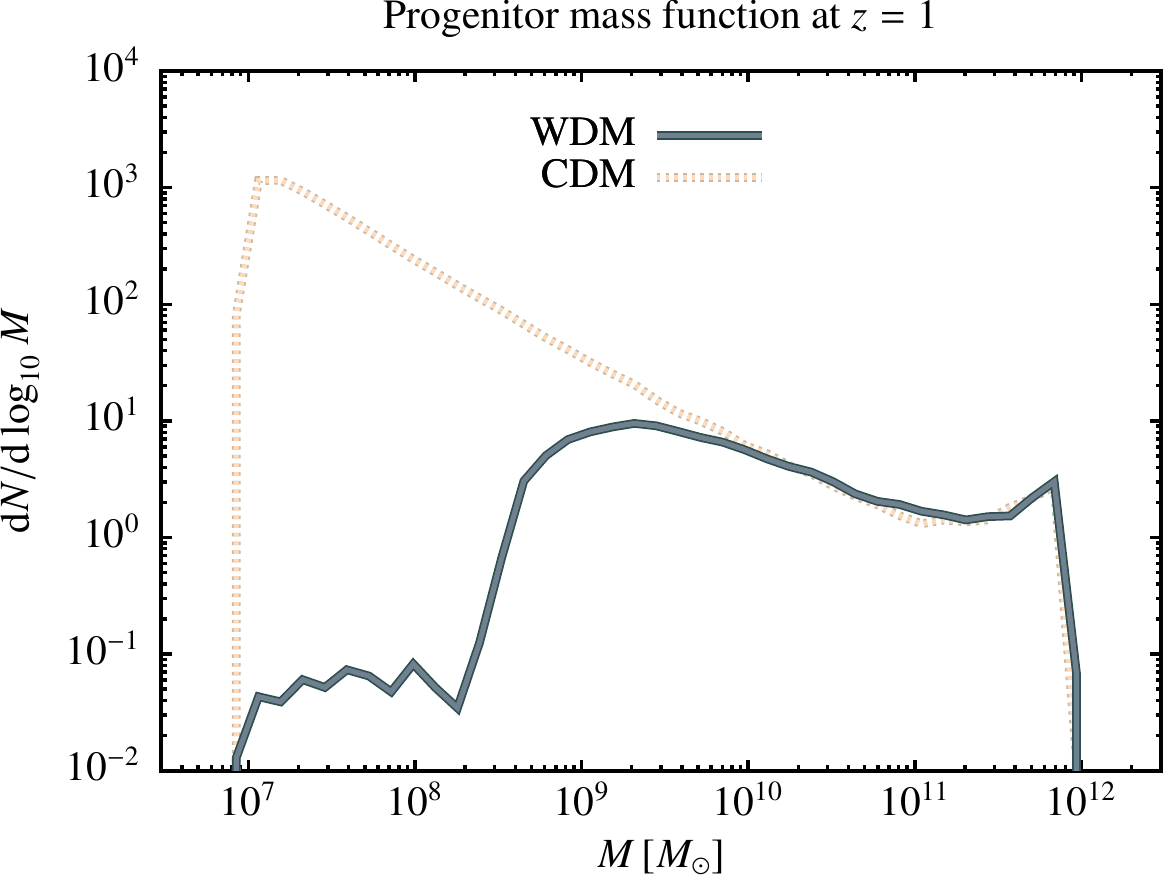} \\
\includegraphics[width=8.5cm]{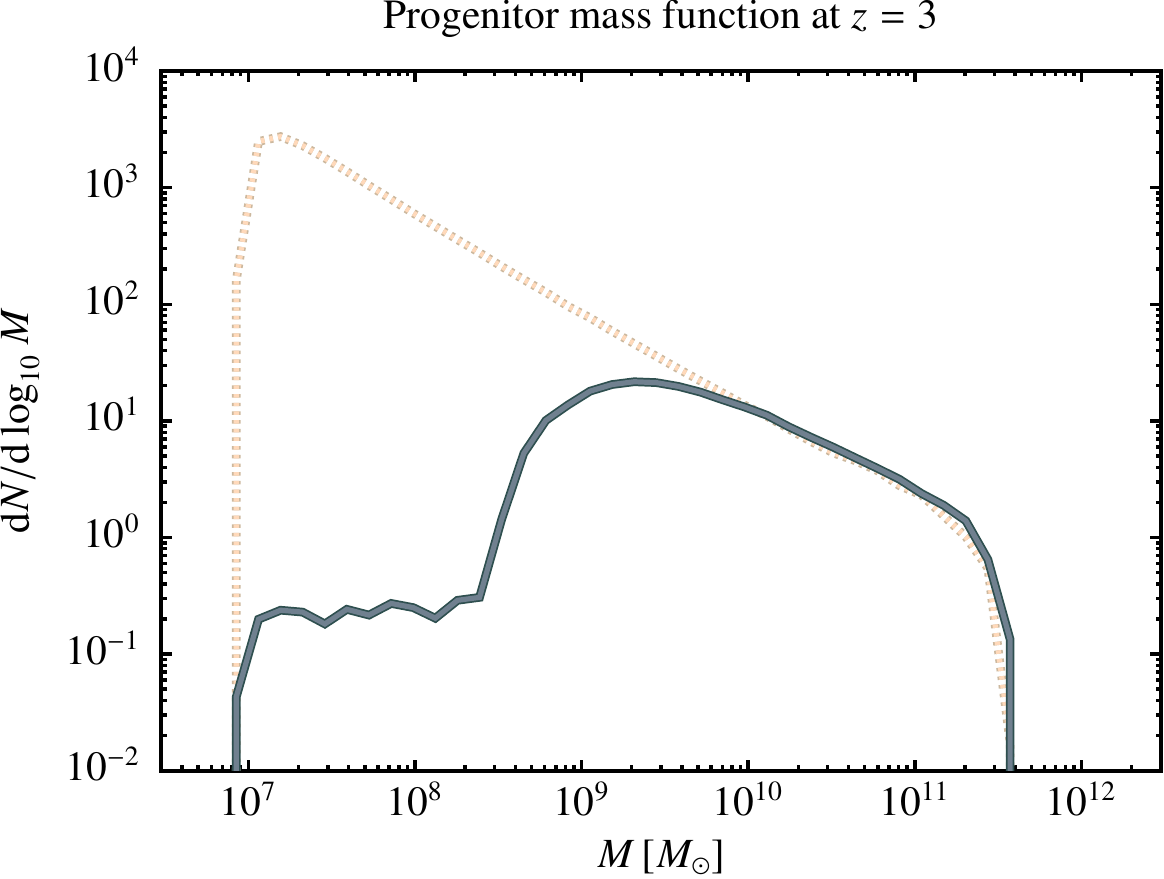} \\
\includegraphics[width=8.5cm]{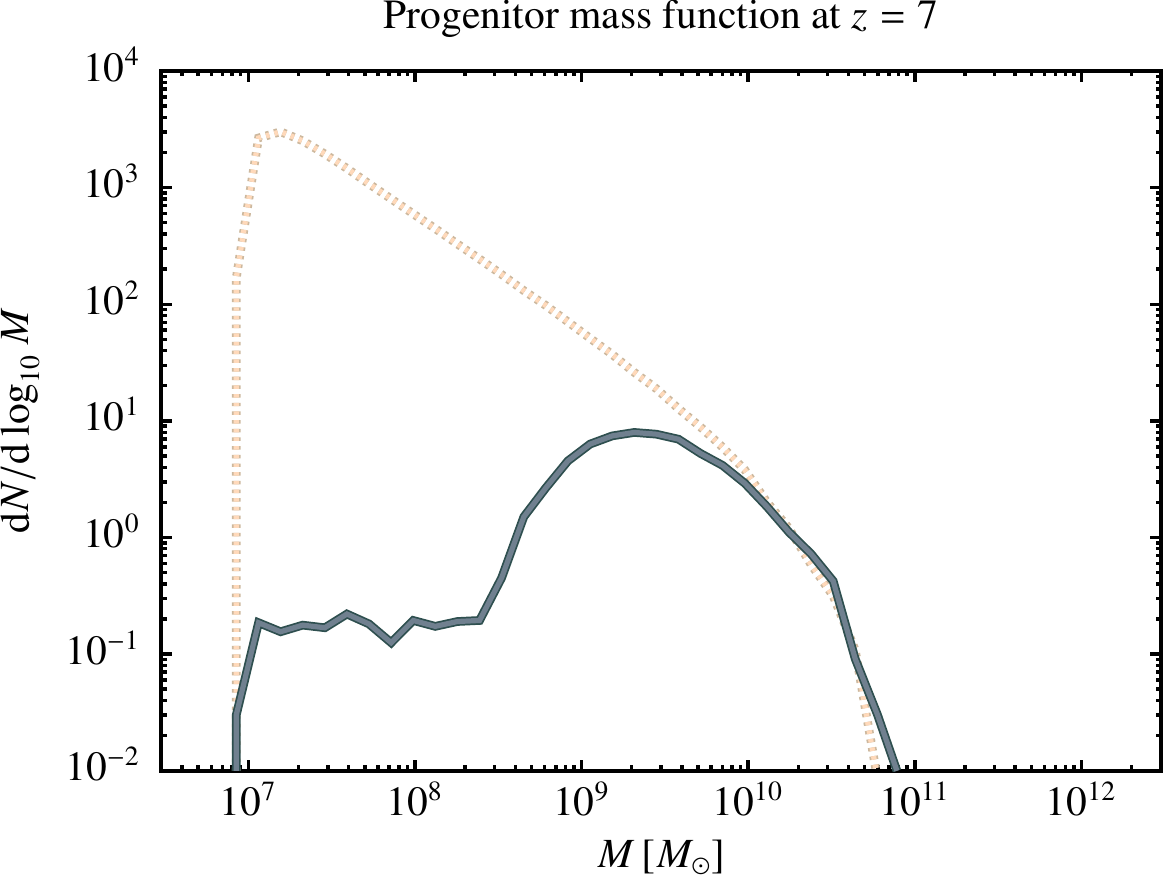}
\end{tabular}
\end{center}
\caption{Progenitor mass functions derived via merger tree construction in \protect\CDM\ and \protect\WDM\ cases for a $10^{12}$~M$_\odot$ halo at $z=0$.}
\label{fig:progenitorMassFunctions}
\end{figure}

Fig.~\ref{fig:progenitorMassFunctions} shows progenitor mass functions at $z=1$, $3$ and $7$. The sharp cut-off at $10^7$~M$_\odot$ (present also in the \CDM\ case) is due to the imposed resolution of our merger trees\footnote{The merger tree resolution is limited only by the available computational time and memory. We choose a resolution of $10^7$~M$_\odot$ in this case to be sufficiently below $M_{\rm s}$ while keeping computing times tractable.} (and so is unphysical). With smooth accretion included, the \WDM\ progenitor mass function closely matches that of \CDM\ above about $3 M_{\rm s}$, but is strongly suppressed below it at lower masses.

Well below the suppression scale a population of progenitor masses much less than $M_{\rm s}$ builds up. These are the result of smooth accretion--the first crossing rate distribution (see Fig.~\ref{fig:firstCrossingWdmVsCdm}) cuts off below $M_{\rm s}$ so there is no way for these halos to arise through branching of the merger tree--which gradually reduces the mass of the lowest mass halos going back in time. The numerical robustness of our model in this regime is discussed in Appendix~\ref{sec:treeNumerics}, in which we also demonstrate that the position of the peak in the progenitor mass function is numerically robust and well determined.

\begin{figure}
\begin{center}
\includegraphics[width=8.5cm]{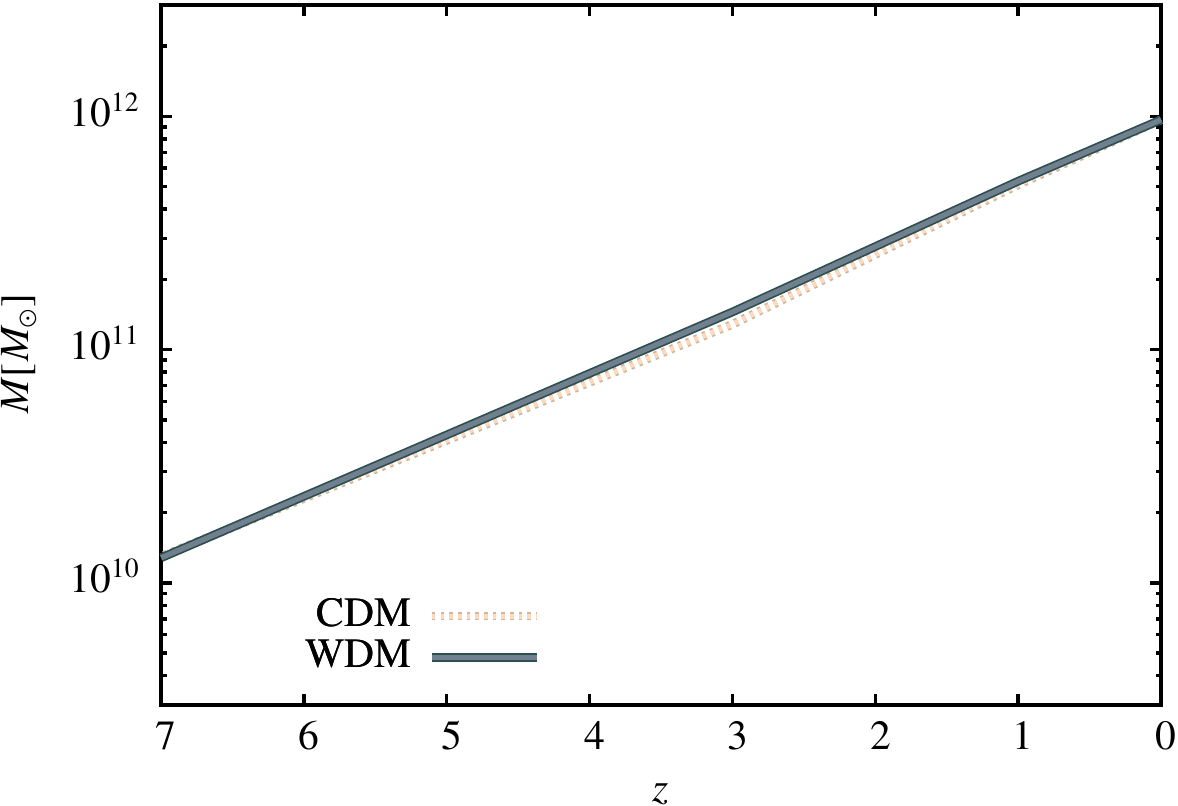}
\end{center}
\caption{Mean mass accretion histories derived via merger tree construction in \protect\CDM\ and \protect\WDM\ cases for a $10^{12}$~M$_\odot$ halo at $z=0$.}
\label{fig:massAccretionHistories}
\end{figure}

Fig.~\ref{fig:massAccretionHistories} shows the corresponding mean mass accretion histories of this halo (i.e. the mean mass of the most massive progenitor at each redshift). There is almost no difference between \CDM\ and \WDM. This is in agreement with the results of \cite{knebe_merger_2002} who found almost no difference in the mass accretion histories of individual halos in \CDM\ and \WDM\ N-body simulations. The two halos presented by \cite{knebe_merger_2002} were significantly more massive (almost $10^{14}$~M$_\odot$) than those considered here, but they report the same conclusion for lower mass halos. \cite{knebe_merger_2002} conclude that the number of mergers that contribute significant mass to the assembly of the halos is unchanged between \CDM\ and \WDM\ case. Our results suggest that this is an accurate conclusion for more massive halos (the assembly of which will be dominated by halos well above $M_{\rm s}$). However, for lower mass halos, which gain a significant fraction of their mass from halos close to $M_{\rm s}$, our results clearly show that smooth accretion plays a crucial role in shaping the mass accretion history of \WDM\ halos--without it, substantial differences from the \CDM\ case would occur.

\section{Discussion}\label{sec:discuss}

We have described algorithms for constructing halo mass functions and merger trees for dark matter halos in \WDM\ universes. Our methods improve upon previous treatments which did not correctly solve the barrier first crossing problem \citep{menci_galaxy_2012} and which used a top-hat filter to compute $\sigma(M)$ resulting in an overestimate of the abundance of low-mass halos. Our results are in excellent agreement with the available N-body simulations. Illustrative results clearly demonstrate that the mass function, and progenitor mass functions of \WDM\ halos are strongly suppressed relative to \CDM\ below about $M_{\rm s}$. Differences between \CDM\ and \WDM\ in coarse-grained statistics, such as the mass accretion history, are small for halos well above the cut-off scale, providing that the accretion of smoothly distributed matter in \WDM\ is accounted for. 

The method that we describe has a single free parameter---the coefficienct $a$ appearing in eqn.~(\ref{eq:kToR}). We have fixed the value of this parameter to match the location of the turnover in the N-body mass function reported by \cite{schneider_non-linear_2012}. This introduces a dependency on one of the simulations to which we compare our model. Nevertheless, as discussed in \S\ref{sec:windowFunction}, we expect that the same value of $a$ will be appropriate for all \WDM\ particle masses of interest. The limited evidence available from our present work (i.e. that $a$ chosen to fit the mass function for $0.25$~keV WDM also successfully matches the progenitor mass functions for $2.20$~keV WDM) certainly supports this claim.

Our approach has the advantage, compared to N-body simulations of \WDM, of not being affected by numerical noise in the particle distribution which leads to the formation of large numbers of artificial low-mass halos \citep{wang_discreteness_2007}, and of being substantially faster to evaluate mass functions and progenitor distributions. Using the techniques developed in this work, the procedure for applying them to dark matter with different phenomenology, or to other physics that modifies the power spectrum or excursion set barrier is straightforward:
\begin{enumerate}
 \item Determine the linear theory power spectrum of density perturbations from the dark matter (or other) physics;
 \item Determine, through analytic calculation or idealized N-body simulations, the critical linear theory overdensity for collapse, which will depend on the (thermal and interaction) physics of the dark matter particle.
\end{enumerate}
Given these two inputs our techniques can be used to determine the resulting halo mass function and merger histories of dark matter halos consistent with the input physics. The accuracy of our methods for phenomenology beyond that exhibited by \WDM\ remains to be tested, but the success in this case leads us to expect that our methods will be generally applicable.

The nature of the dark matter distribution on small mass scales will be investigated by future lensing programs \citep{keeton_new_2009,vegetti_bayesian_2009}. The techniques described in this work will allow detailed statistical predictions to be made for the expected numbers and masses of dark matter substructure as a function of dark matter particle properties. 

To make accurate predictions for the dark matter subhalo distribution we have addressed only the first part of the problem, namely halo formation and merging. The second part, halo destruction by tidal forces must also be addressed. We plan to explore this process using the methods of \citeauthor{benson_effects_2002-1}~(\citeyear{benson_effects_2002-1}; see also \citealt{taylor_evolution_2004}), together with prescriptions for the internal structure of \WDM\ halos (which will, of course, differ from that of \CDM\ halos; \citealt{maccio_inner_2012,maccio_cores_2012}).

The methods described in \S\ref{sec:methods} have been implemented within the open source semi-analytic galaxy formation code, {\sc galacticus} \citep{benson_galacticus:_2012}. All results presented in this work were generated using {\sc galacticus} {\tt v0.9.1 r903}. A control files and scripts to generate all results presented in this paper using {\sc galacticus} can be found at  \href{http://users.obs.carnegiescience.edu/abenson/galacticus/parameters/dmMergingBeyondCDM1.tar.bz2}{{\tt http://users.obs.carnegiescience.edu/abenson/\newline galacticus/parameters/dmMergingBeyondCDM1.tar.bz2}}.

\section*{Acknowledgments}

We thank Rennan Barkana for supplying results from his calculations of halo collapse in \WDM\ universes, and Annika Peter for invaluable discussions and comments on an earlier draft of this paper. The work of LAM was carried out at Jet Propulsion Laboratory, California Institute of Technology, under a contract with NASA. LAM acknowledges support by the NASA ATFP program.

\bibliographystyle{mn2e}
\bibliography{bensonWDM}

\appendix
\onecolumn

\section{Numerical Method}\label{app:numericalMethod}

In this section, we describe our numerical method for solving the excursion set barrier first crossing problem (eqn.~\ref{eq:OldExcursionMethod}). In the absence of a barrier, $P(\delta,S)$ would be equal to $P_0(\delta,S)$ which is simply a Gaussian distribution with variance $S$:
\begin{equation}
  P_0(\delta,S) = \frac{1}{\sqrt{2 \pi S}} \exp\left(-{\delta^2 \over 2 S}\right).
  \label{eq:Gaussian}
\end{equation}
Since the barrier absorbs any random walks which cross is at smaller $S$, the actual $P(\delta,S)$ must therefore be given by:
\begin{equation}
   P(\delta,S) = P_0(\delta,S) - \int_{0}^{S} f(S^\prime) P_0[\delta - B(S^\prime),S - S^\prime]{\rm d}S^\prime .
 \label{eq:Displaced}
\end{equation}
The second term on the right hand side of eqn.~(\ref{eq:Displaced}) represents the distribution of random trajectories originating from the point $(S,B(S))$. The integral therefore gives the fraction of trajectories which crossed the barrier at $S<S^\prime$ and which can now be found at $(S,\delta)$.

Using this result, we can rewrite eqn.~(\ref{eq:OldExcursionMethod}):
\begin{equation}
  1 = \int_0^S f(S^\prime){\rm d}S^\prime + \int_{-\infty}^{B(S)} \left[ P_0(\delta,S) -  \int_{0}^{S} f(S^\prime) P_0(\delta - B(S^\prime),S - S^\prime){\rm d}S^\prime )\right] {\rm d} \delta ,
\end{equation}
in general and, for the Gaussian distribution of eqn.~(\ref{eq:Gaussian}):
\begin{equation}
  1 =  \int_0^S f(S^\prime){\rm d}S^\prime + \int_{-\infty}^{B(S)} \left[ \frac{1}{\sqrt{2 \pi S}} \exp\left(-\frac{\delta^2}{2 S}\right) -  \int_{0}^{S} f(S^\prime) \frac{1}{\sqrt{2 \pi (S-S^\prime)}} \exp\left(-\frac{[\delta - B(S^\prime)]^2}{2 (S-S^\prime)}\right){\rm d}S^\prime \right] {\rm d} \delta .
\end{equation}
The integral over ${\rm d}\delta$ can be carried out analytically to give:
\begin{equation}
 1 =  \int_0^S f(S^\prime){\rm d}S^\prime+  \hbox{erf}\left[\frac{B(S)}{\sqrt{2S}}\right]  -  \int_{0}^{S}  f(S^\prime)  \hbox{erf}\left[\frac{B(S) - B(S^\prime)}{\sqrt{2 (S-S^\prime)}}\right] {\rm d}S^{\prime\prime}.
\label{eq:NewExcursionMethod}
\end{equation}
We now discretize eqn.~(\ref{eq:NewExcursionMethod}). Specifically, we divide the $S$ space into $N$ intervals defined by the points:
\begin{equation}
  S_i = \left\{ \begin{array}{ll}
                 0 & \hbox{if } i=0 \\
                 \sum_0^{i-1} \Delta S_i & \hbox{if } i > 1.
                \end{array}
        \right.
\end{equation}
Note that $f(0)=0$ by definition, so $f(S_0)=0$ always. We choose $\Delta S_i = S_{\rm max}/N$ (i.e. uniform spacing in $S$) when computing first crossing distributions, and $\Delta S_i \propto S_i$ (i.e. uniform spacing in $\log(S)$) when computing first crossing rates.

Discretizing the integrals in eqn.~(\ref{eq:NewExcursionMethod}) gives:
\begin{equation} \label{eq:Des1}
 \int_0^{S_j} f(S^\prime)\d S^\prime = \sum_{i=0}^{j-1} \frac{f(S_i) + f(S_{i+1})}{2} \Delta S_i
\end{equation}
and:
\begin{equation} \label{eq:Des2}
 \int_{0}^{S_j}  f(S^\prime)  \hbox{erf}\left[\frac{B(S) - B(S^\prime)}{\sqrt{2 (S-S^\prime)}}\right] \d S^\prime = \sum_{i=0}^{j-1} \frac{1}{2} \left(f(S_i)  \hbox{erf}\left[\frac{B(S_j) - B(S_i)}{\sqrt{2 (S_j-S_i)}}\right] + f(S_{i+1})  \hbox{erf}\left[\frac{B(S_j) - B(S_{i+1})}{\sqrt{2 (S_j-S_{i+1})}}\right] \right) \Delta S_i.
\end{equation}
We can now rewrite eqn.~(\ref{eq:NewExcursionMethod}) in discretized form:
\begin{equation} \label{eq:DesFinal1}
 1 = \sum_{i=0}^{j-1} \frac{f(S_i) + f(S_{i+1})}{2} \Delta S_i  +  \hbox{erf}\left[\frac{B(S_j)}{\sqrt{2S_j}}\right] - \frac{1}{2} \sum_{i=0}^{j-1} \left( f(S_i)  \hbox{erf}\left[\frac{B(S_j) - B(S_i)}{\sqrt{2 (S_j-S_i)}}\right] + f(S_{i+1})  \hbox{erf}\left[\frac{B(S_j) - B(S_{i+1})}{\sqrt{2 (S_j-S_{i+1})}}\right]  \right)  \Delta S_i.
\end{equation}
Solving eqn.~(\ref{eq:DesFinal1}) for $f(S_j)$:
\begin{eqnarray} \label{eq:DesFinal11}
 \left( \frac{1}{2} - \frac{1}{2} \hbox{erf}\left[\frac{B(S_j) - B(S_j)}{\sqrt{2 (S_j-S_j)}}\right] \right) \Delta S_{j-1} f(S_j) &=& 1 - \sum_{i=0}^{j-2} \frac{f(S_i) + f(S_{i+1})}{2} \Delta S_i - \frac{f(S_{j-1})}{2} \Delta S_{j-1} -  \hbox{erf}\left\{\frac{B(S_j)}{\sqrt{2S_j}}\right\}   \nonumber\\
& & + \frac{1}{2} \sum_{i=0}^{j-2} \left( f(S_i)  \hbox{erf}\left\{\frac{[B(S_j) - B(S_i)]}{\sqrt{2 (S_j-S_i)}}\right\} + f(S_{i+1})  \hbox{erf}\left\{\frac{[B(S_j) - B(S_{i+1})]}{\sqrt{2 (S_j-S_{i+1})}}\right\} \right)\Delta S_i  \nonumber \\
 & & + \frac{1}{2} f(S_{j-1})  \hbox{erf}\left\{\frac{[B(S_j) - B(S_{j-1})]}{\sqrt{2 (S_j-S_{j-1})}}\right\} \Delta S_{j-1}.
\end{eqnarray}
For all barriers that we consider:
\begin{equation} 
\hbox{erf}\left[\frac{B(S_j) - B(S_j)}{\sqrt{2 (S_j-S_j)}}\right] = 0.
\end{equation}
We can then simplify eqn.~(\ref{eq:DesFinal11}):
\begin{eqnarray} \label{eq:DesFinal2}
   f(S_j) &=& {2 \over \Delta S_{j-1}}\left[1 - \sum_{i=0}^{j-2} \frac{f(S_i) + f(S_{i+1})}{2} \Delta S_i - \frac{f(S_{j-1})}{2} \Delta S_{j-1} -  \hbox{erf}\left\{\frac{B(S_j)}{\sqrt{2S_j}}\right\} \right.  \nonumber\\
& & + \frac{1}{2} \sum_{i=0}^{j-2} \left( f(S_i)  \hbox{erf}\left\{\frac{[B(S_j) - B(S_i)]}{\sqrt{2 (S_j-S_i)}}\right\} + f(S_{i+1})  \hbox{erf}\left\{\frac{[B(S_j) - B(S_{i+1})]}{\sqrt{2 (S_j-S_{i+1})}}\right\} \right)\Delta S_i  \nonumber \\
 & & \left. + \frac{1}{2} f(S_{j-1})  \hbox{erf}\left\{\frac{[B(S_j) - B(S_{j-1})]}{\sqrt{2 (S_j-S_{j-1})}}\right\} \Delta S_{j-1}\right].
\end{eqnarray}
Consolidating terms in the summations:
\begin{equation} \label{eq:DesFinal2a}
   f(S_j) = {2 \over \Delta S_{j-1}}\left[1 -  \hbox{erf}\left\{\frac{B(S_j)}{\sqrt{2S_j}}\right\} - \sum_{i=0}^{j-1} \left\{ 1-\hbox{erf}\left[\frac{B(S_j) - B(S_i)}{\sqrt{2 (S_j-S_i)}}\right] \right\} f(S_i) {\Delta S_{i-1} + \Delta S_i \over 2} \right].
\end{equation}
In the case of constant $\Delta S_i(=\Delta S)$ this can be simplified further:
\begin{equation} \label{eq:DesFinal3}
   f(S_j) = {2 \over \Delta S}\left[1 - \hbox{erf}\left\{\frac{B(S_j)}{\sqrt{2S_j}}\right\}\right] - 2 \sum_{i=0}^{j-1} \left\{1- \hbox{erf}\left[\frac{B(S_j) - B(S_i)}{\sqrt{2 (S_j-S_i)}}\right] \right\} f(S_i).
\end{equation}

In either case (i.e. eqns.~\ref{eq:DesFinal2a} and \ref{eq:DesFinal3}) solution proceeds recursively: $f(S_0)=0$ by definition, $f(S_1)$ depends only on the known barrier and $f(S_0)$, $f(S_j)$ depends only on the known barrier and $f(S_{<j})$.

\twocolumn
\section{Numerical Tests}\label{app:numerics}

In this Appendix we examine the numerical accuracy and robustness of our methods.

\subsection{First Crossing Probability Solutions}\label{sec:firstCrossingNumerics}

\begin{figure}
\begin{center}
\includegraphics[width=8.5cm]{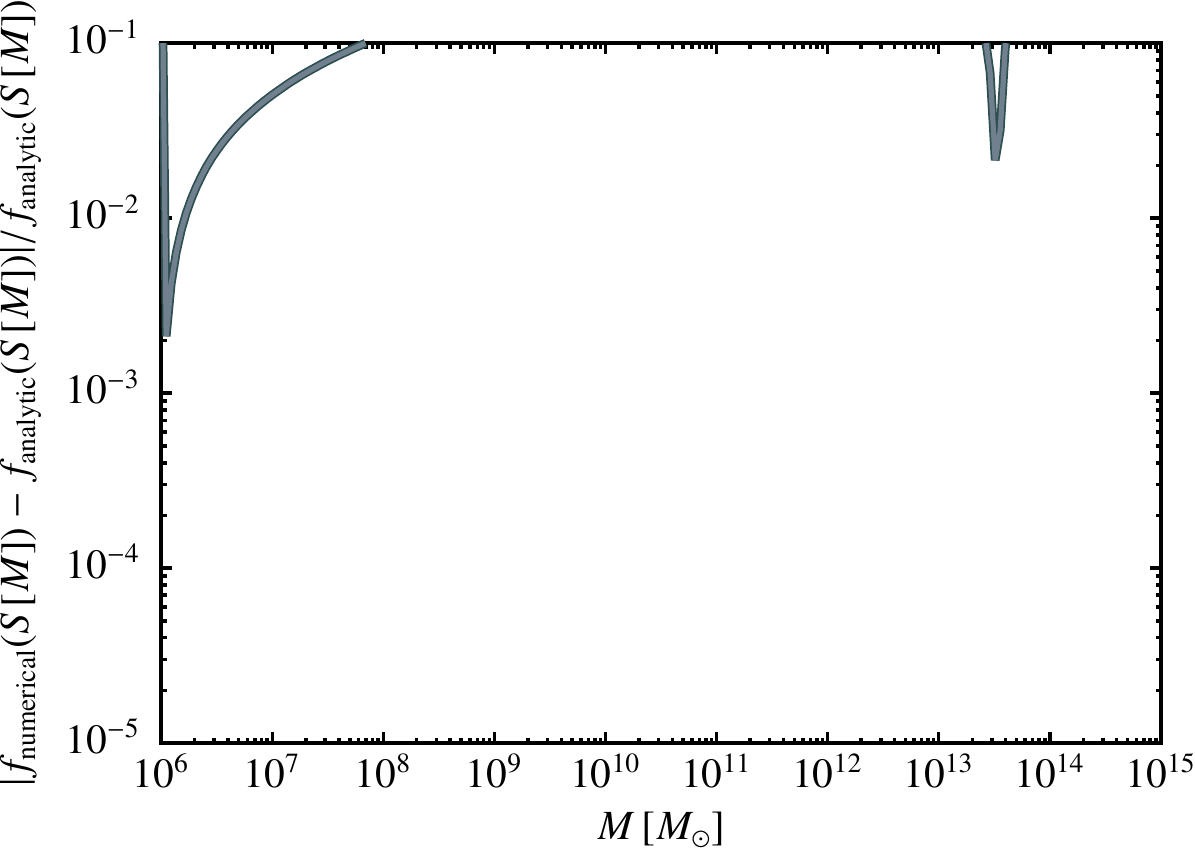}
\end{center}
\caption{The absolute fractional error in the excursion set first crossing distribution function, $f(S)$, computed using our numerical method compared to the analytic solution for a constant barrier.}
\label{fig:numericalVsAnalyticSolver}
\end{figure}

To test the accuracy of our numerical solver for the first crossing distribution we compare its results to the known analytic solution for a constant barrier. Specifically, we consider the \CDM\ case with no remapping of the barrier such that $B(S)=\delta_{\rm c}=$constant and
\begin{equation}
 f(S) = {\delta_{\rm c} \over S\sqrt{2\pi S}} \exp\left[-{\delta^2_{\rm c}\over 2 S}\right].
 \label{eq:LinearBarrierSolution}
\end{equation}
Fig.~\ref{fig:numericalVsAnalyticSolver} shows the fractional difference between results obtained using our numerical method and the analytic solution in this case. The numerical error is small except for at the highest masses (smallest variances) where the discreteness in our grid becomes an issue and the error reaches a few percent. The accuracy achieved is sufficient for the present work (where we are mostly concerned with low-mass systems) and is easily improved by adopting a smaller value of $\Delta S$.

\subsection{First Crossing Rate Solutions}\label{sec:rateNumerics}

\begin{figure}
\begin{center}
\includegraphics[width=8.5cm]{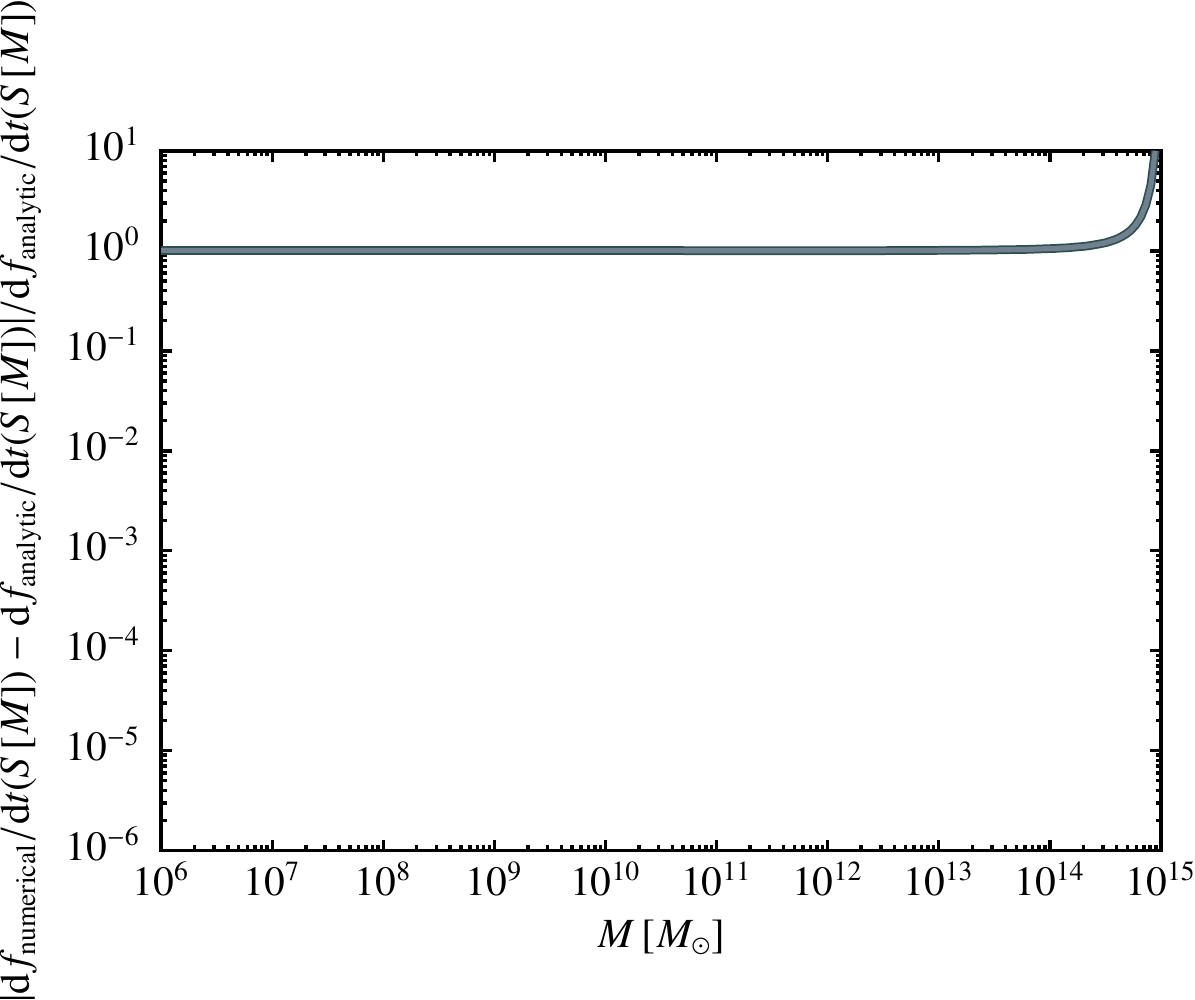}
\end{center}
\caption{The absolute fractional error in the excursion set first crossing rate function, ${\rm d} f(S)/{\rm d} t$, computed using our numerical method compared to the analytic solution for a constant barrier.}
\label{fig:numericalVsAnalyticRateSolver}
\end{figure}

Once again, to test the accuracy of our numerical solver in the case of computing first crossing rates we compare its results to the analytic solution for a constant barrier. Adopting the same constant barrier model as in Appendix~\ref{sec:firstCrossingNumerics} the crossing rate is
\begin{equation}
 {{\rm d} f \over {\rm d} t} = {1 \over \sqrt{2 \pi}} {1 \over (S-S_0)^{3/2}} {{\rm d} \delta_{\rm c} \over {\rm d} t},
\end{equation}
where $S_0$ is the variance corresponding to the mass of the final halo (i.e. the halo formed through the merger event). Fig.~\ref{fig:numericalVsAnalyticRateSolver} shows the fractional error in the numerically derived merger rate compared to this analytic solution for $\epsilon=0.01$. The fractional error is constant across the range of masses shown and is limited by the value of $\epsilon$ chosen.

\subsection{Testing the Parkinson-Cole-Helly Algorithm on Smaller Mass Scales}\label{sec:pchNumerics}

\begin{figure*}
\begin{center}
\begin{tabular}{cc}
\includegraphics[width=8.5cm]{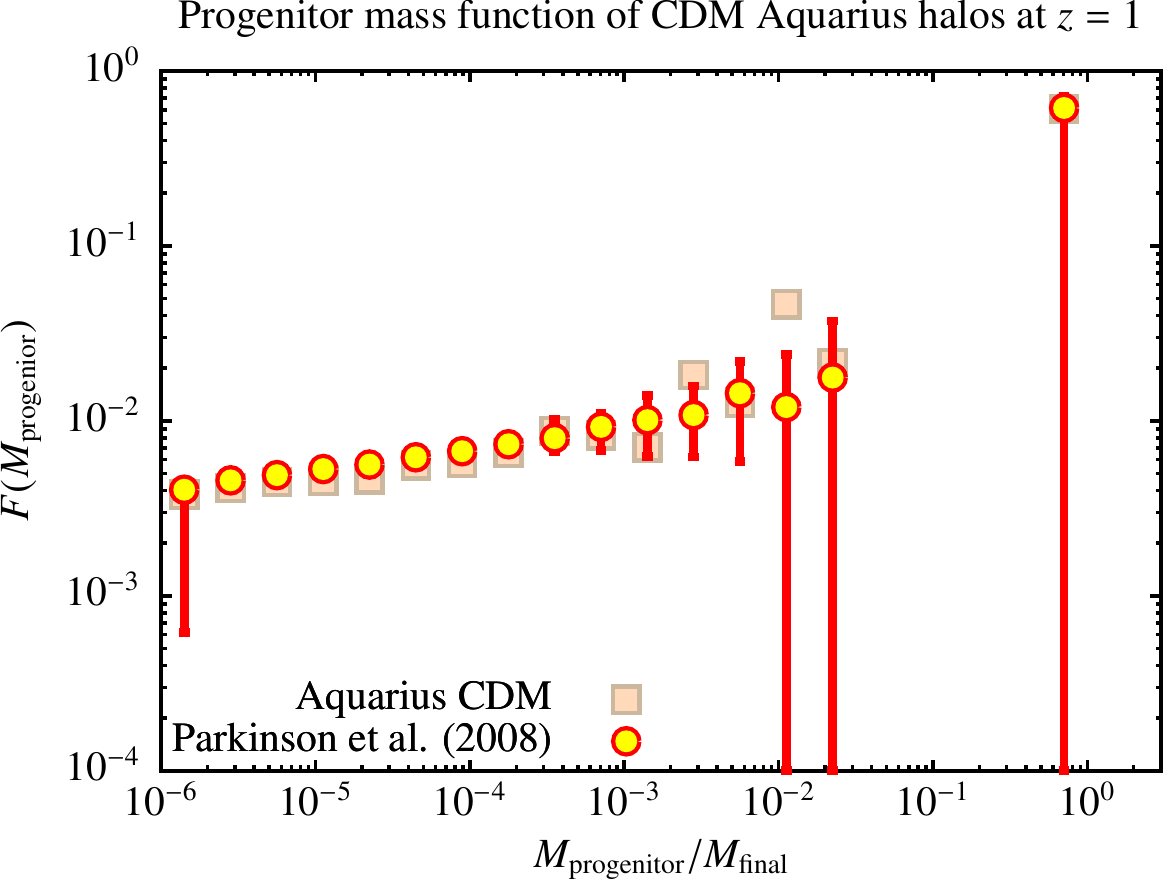} &
\includegraphics[width=8.5cm]{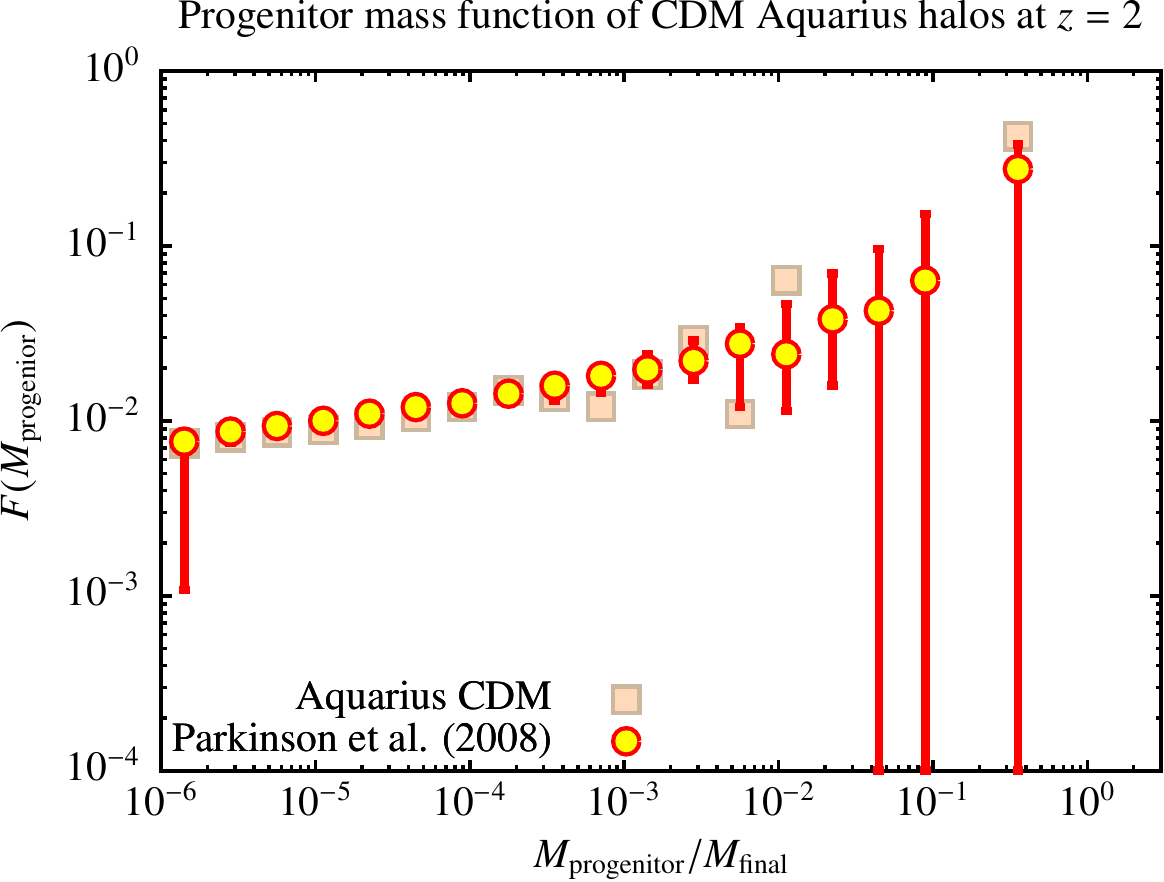}
\end{tabular}
\end{center}
\caption{Progenitor mass functions of the Aquarius \protect\CDM\ dark matter halos \protect\citep{springel_aquarius_2008} at $z=1$ and 2 (histograms) compared with the predictions of the \protect\cite{parkinson_generating_2008} algorithm. Each panel shows the fraction of the halo's mass contributed by progenitors in each mass bin. Masses are shown as a fraction of the final halo mass. Error bars on the predictions from the \protect\cite{parkinson_generating_2008} algorithm indicate the 20$^{\rm th}$ and $80^{\rm th}$ percentiles of the distribution of progenitor mass functions. While this algorithm was calibrated on the much higher mass halos found in the Millennium Simulation \protect\citep{springel_simulations_2005} it can be seen to work extremely well at these lower masses also.}
\label{fig:PCH}
\end{figure*}

The \cite{parkinson_generating_2008} empirical modification to the merger tree branching rate was calibrated against N-body merger trees drawn from the Millennium Simulation. As such, it has been tested for masses above approximately $10^{10}{\rm M}_\odot$. Here we employ this same modification for much lower masses. Fig.~\ref{fig:PCH} compares progenitor mass functions generated by the \cite{parkinson_generating_2008} empirical modification with those extracted from the Aquarius \CDM\ simulations of \cite{springel_aquarius_2008} which resolve halos of masses $10^6{\rm M}_\odot$. It can be seen that the \cite{parkinson_generating_2008} empirical modification performs very well (with unchanged parameter values) for these much lower mass halos in the \CDM\ case also.

\subsection{Merger Tree Accuracy and Smooth Accretion}\label{sec:treeNumerics}

To test the convergence of our merger trees with respect to the parameter $\epsilon$ used in our numerical determination of the first crossing rate distribution (see \S\ref{sec:mergerRates}) we compute progenitor mass functions of $10^{12}$M$_\odot$ halos for $\epsilon=0.010$, $0.003$, and $0.001$. Additionally, we perform these calculations both with and without the smooth accretion term of \S\ref{sec:smoothAccretion}.

\begin{figure}
\begin{center}
\begin{tabular}{c}
\includegraphics[width=8.5cm]{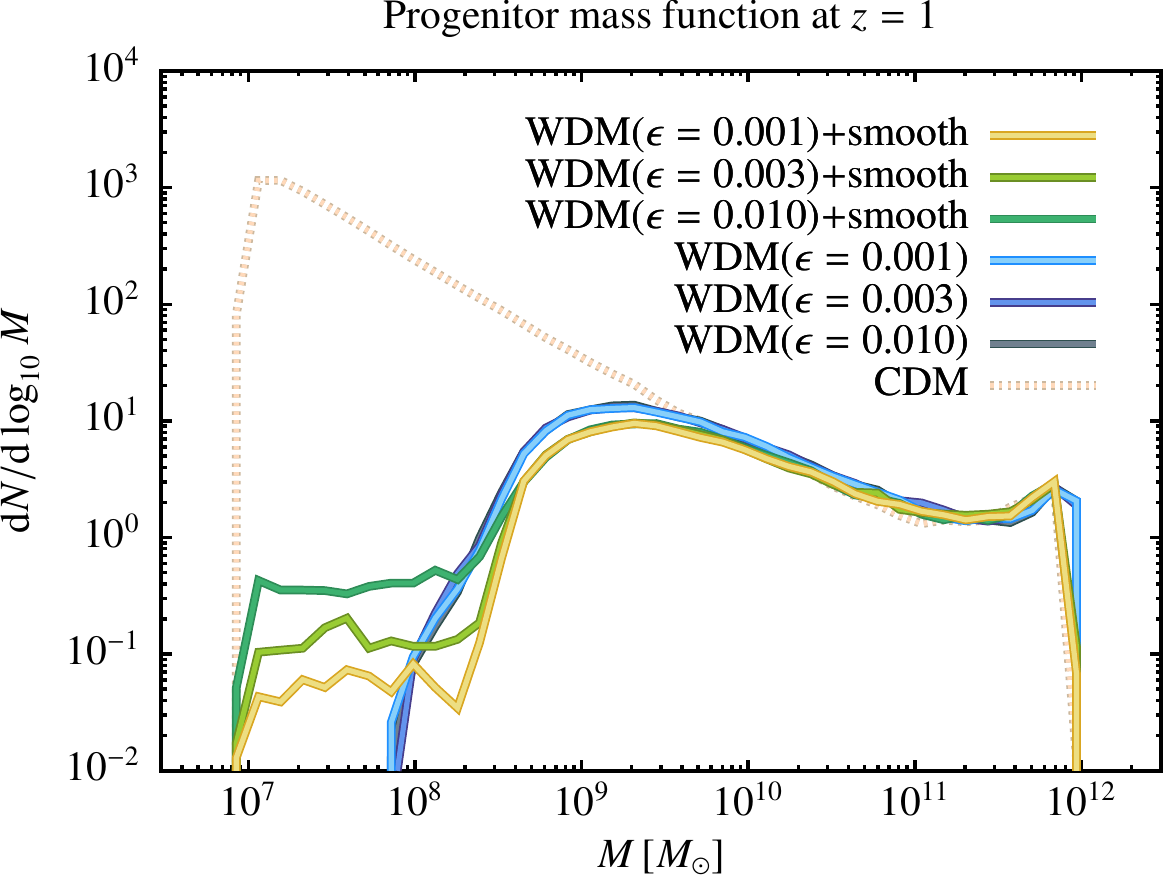} \\
\includegraphics[width=8.5cm]{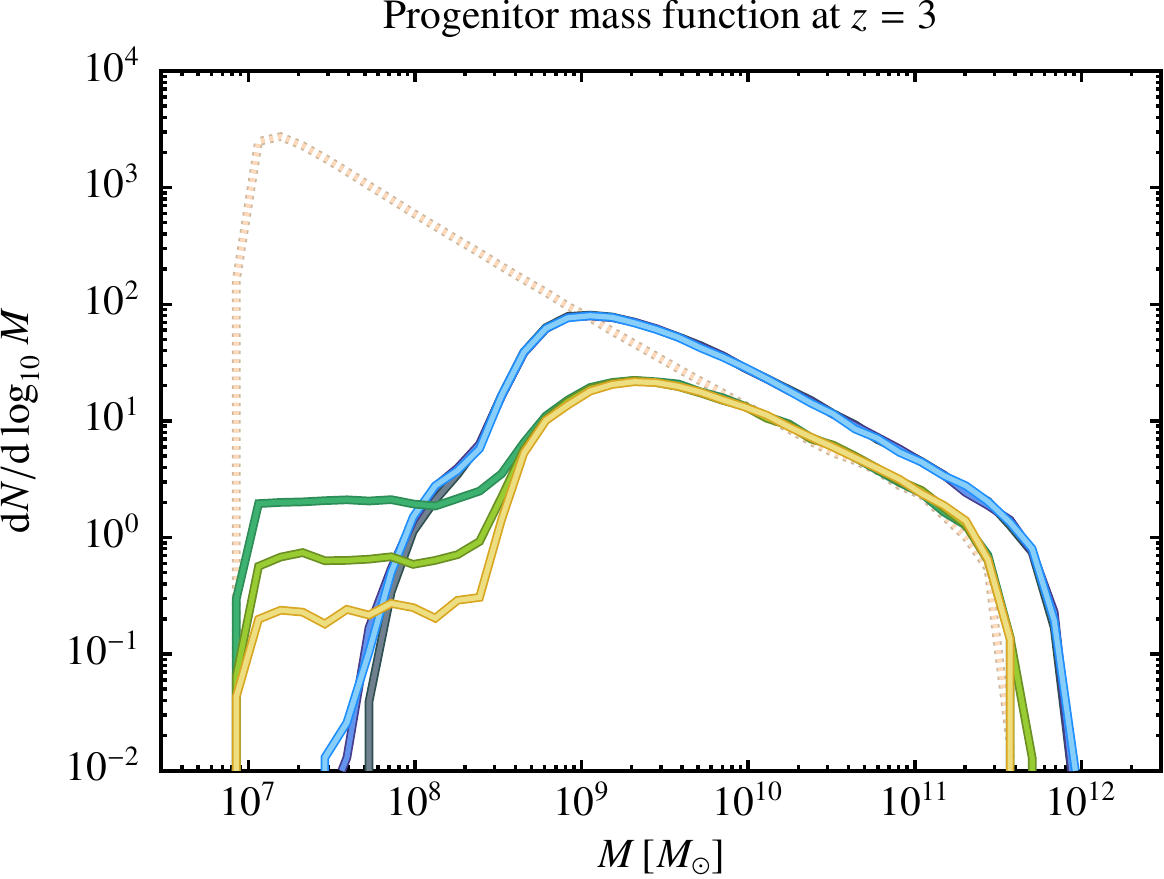} \\
\includegraphics[width=8.5cm]{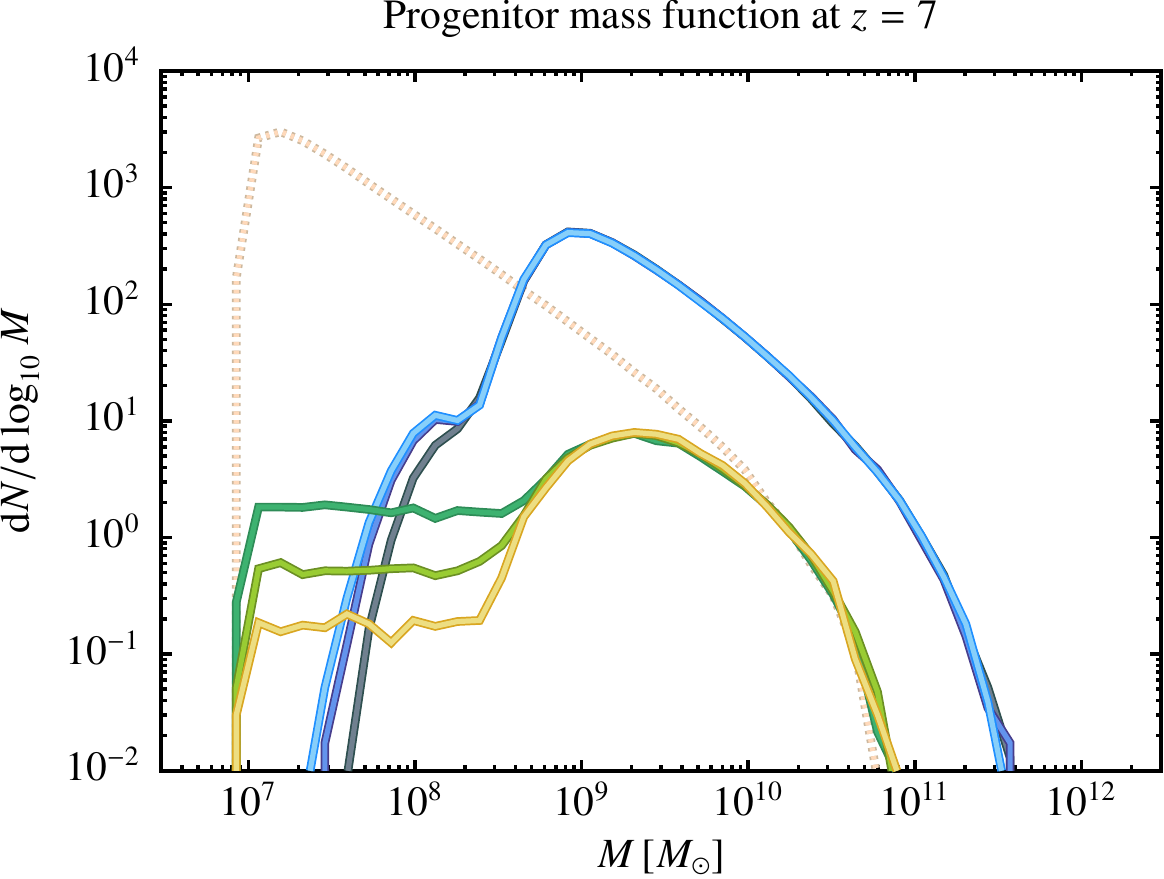}
\end{tabular}
\end{center}
\caption{Progenitor mass functions derived via merger tree construction in \protect\CDM\ and \protect\WDM\ cases for a $10^{12}$~M$_\odot$ halo at $z=0$. For the \protect\WDM\ case, results are shown for different values of $\epsilon$ and also for cases with and without the accretion of smoothly distributed matter.}
\label{fig:progenitorMassFunctionsNumerics}
\end{figure}

Fig.~\ref{fig:progenitorMassFunctionsNumerics} shows progenitor mass functions at $z=1$, $3$ and $7$. When smooth accretion is included, the \WDM\ progenitor mass function closely matches that of \CDM\ above about $3 M_{\rm s}$, but is strongly suppressed below it at lower masses. If smooth accretion is ignored the \WDM\ progenitor mass function evolves much more slowly and diverges from the \CDM\ case even at the highest masses. Ignoring this smooth accretion leads to significantly biased results.

It can clearly be seen that our \WDM\ results are converged with respect to the $\epsilon$ parameter except at masses well below the suppression scale in the progenitor mass function in cases where smooth accretion is included. Here, a population of progenitor masses much less than $M_{\rm s}$ builds up. These are the result of smooth accretion--the first crossing rate distribution (see Fig.~\ref{fig:firstCrossingWdmVsCdm}) cuts off below $M_{\rm s}$ so there is no way for these halos to arise through branching of the merger tree--which gradually reduces the mass of the lowest mass halos going back in time.

Our numerical determination of the first crossing rate function is currently not robust in its determination of smooth accretion rates for these lowest mass (highest variance) halos where the excursion set barrier is a very rapidly changing function of variance and our discretization of $S$ used to obtain a numerical solution inevitably does a poor job of resolving the barrier. This could of course be mitigated by using a yet smaller value of $\epsilon$ (requiring substantially increased precision in the numerical solutions) or a finer grid in $S$ (requiring both increased precision and substantially more computing time). Nevertheless, the position of the peak in the progenitor mass function is well determined, and the differences in the abundance of low mass progenitors have negligible effect on the mean mass accretion histories of halos considered in this work.

\end{document}